\PassOptionsToPackage{unicode}{hyperref}
\PassOptionsToPackage{hyphens}{url}
\PassOptionsToPackage{dvipsnames,svgnames,x11names}{xcolor}
\documentclass[
  12pt]{article}

\usepackage{amsmath,amssymb}
\usepackage{iftex}
\ifPDFTeX
  \usepackage[T1]{fontenc}
  \usepackage[utf8]{inputenc}
  \usepackage{textcomp} 
\else 
  \usepackage{unicode-math}
  \defaultfontfeatures{Scale=MatchLowercase}
  \defaultfontfeatures[\rmfamily]{Ligatures=TeX,Scale=1}
\fi
\usepackage{lmodern}
\ifPDFTeX\else  
\fi
\IfFileExists{upquote.sty}{\usepackage{upquote}}{}
\IfFileExists{microtype.sty}{
  \usepackage[]{microtype}
  \UseMicrotypeSet[protrusion]{basicmath} 
}{}
\makeatletter
\@ifundefined{KOMAClassName}{
  \IfFileExists{parskip.sty}{%
    \usepackage{parskip}
  }{
    \setlength{\parindent}{0pt}
    \setlength{\parskip}{6pt plus 2pt minus 1pt}}
}{
  \KOMAoptions{parskip=half}}
\makeatother
\usepackage{xcolor}
\makeatletter
\ifx\paragraph\undefined\else
  \let\oldparagraph\paragraph
  \renewcommand{\paragraph}{
    \@ifstar
      \xxxParagraphStar
      \xxxParagraphNoStar
  }
  \newcommand{\xxxParagraphStar}[1]{\oldparagraph*{#1}\mbox{}}
  \newcommand{\xxxParagraphNoStar}[1]{\oldparagraph{#1}\mbox{}}
\fi
\ifx\subparagraph\undefined\else
  \let\oldsubparagraph\subparagraph
  \renewcommand{\subparagraph}{
    \@ifstar
      \xxxSubParagraphStar
      \xxxSubParagraphNoStar
  }
  \newcommand{\xxxSubParagraphStar}[1]{\oldsubparagraph*{#1}\mbox{}}
  \newcommand{\xxxSubParagraphNoStar}[1]{\oldsubparagraph{#1}\mbox{}}
\fi
\makeatother

\usepackage{longtable,booktabs,array}
\usepackage{calc} 
\usepackage{etoolbox}
\makeatletter
\patchcmd\longtable{\par}{\if@noskipsec\mbox{}\fi\par}{}{}
\makeatother
\IfFileExists{footnotehyper.sty}{\usepackage{footnotehyper}}{\usepackage{footnote}}
\makesavenoteenv{longtable}
\usepackage{graphicx}
\makeatletter
\def\maxwidth{\ifdim\Gin@nat@width>\linewidth\linewidth\else\Gin@nat@width\fi}
\def\maxheight{\ifdim\Gin@nat@height>\textheight\textheight\else\Gin@nat@height\fi}
\makeatother
\setkeys{Gin}{width=\maxwidth,height=\maxheight,keepaspectratio}
\makeatletter
\def\fps@figure{htbp}
\makeatother

\makeatletter
\@ifpackageloaded{caption}{}{\usepackage{caption}}
\AtBeginDocument{%
\ifdefined\contentsname
  \renewcommand*\contentsname{Table of contents}
\else
  \newcommand\contentsname{Table of contents}
\fi
\ifdefined\listfigurename
  \renewcommand*\listfigurename{List of Figures}
\else
  \newcommand\listfigurename{List of Figures}
\fi
\ifdefined\listtablename
  \renewcommand*\listtablename{List of Tables}
\else
  \newcommand\listtablename{List of Tables}
\fi
\ifdefined\figurename
  \renewcommand*\figurename{Figure}
\else
  \newcommand\figurename{Figure}
\fi
\ifdefined\tablename
  \renewcommand*\tablename{Table}
\else
  \newcommand\tablename{Table}
\fi
}
\@ifpackageloaded{float}{}{\usepackage{float}}
\floatstyle{ruled}
\@ifundefined{c@chapter}{\newfloat{codelisting}{h}{lop}}{\newfloat{codelisting}{h}{lop}[chapter]}
\floatname{codelisting}{Listing}

\makeatother
\makeatletter
\@ifpackageloaded{caption}{}{\usepackage{caption}}
\@ifpackageloaded{subcaption}{}{\usepackage{subcaption}}
\makeatother

\ifLuaTeX
  \usepackage{selnolig}  
\fi
\usepackage[]{natbib}
\usepackage{bookmark}

\IfFileExists{xurl.sty}{\usepackage{xurl}}{} 
\hypersetup{
  pdftitle={Title},
  pdfauthor={Author 1; Author 2},
  pdfkeywords={3 to 6 keywords, that do not appear in the title},
  colorlinks=true,
  linkcolor={blue},
  filecolor={Maroon},
  citecolor={Blue},
  urlcolor={Blue},
  pdfcreator={LaTeX via pandoc}}

\newcommand{\yth}{\boldsymbol{\theta},\boldsymbol{y}}
\newcommand*\diff{\mathop{}\!\mathrm{d}}

\definecolor{step1blue}{HTML}{7AA7F1}
\definecolor{step2blue}{HTML}{648FFF}
\definecolor{step3purple}{HTML}{785EF0}
\definecolor{step4pink}{HTML}{DC267F}
\definecolor{step5pink}{HTML}{FE6100}
\definecolor{step6orange}{HTML}{FFB000}
\DeclareMathOperator*{\argmax}{arg\,max}
\DeclareMathOperator*{\argmin}{arg\,min}

\usepackage[ruled,vlined]{algorithm2e}
\usepackage{mathdots}
\usepackage{tikz}
\usetikzlibrary{arrows.meta, positioning, shapes.geometric, calc}

\newcommand{\anon}{1}

\begin{document}

\def\spacingset#1{\renewcommand{\baselinestretch}%
{#1}\small\normalsize} \spacingset{1}


\if1\anon
{
  \title{\bf Neural Simulation-Based Inference for space-time Gaussian processes}
\author{
    Alexandre Loret$^{1}$, 
    Lucia Clarotto$^{2}$, 
    Nicolas Desassis$^{1}$, 
    Thomas Romary$^{1}$ \\[0.2cm]
    $^{1}$STIM, Mines Paris - PSL \\
    $^{2}$MIA Paris-Saclay, AgroParisTech, INRAE, Université Paris-Saclay \\[0.2cm]
}
\maketitle}\fi

\if0\anon
{
  \bigskip
  \bigskip
  \bigskip
  \begin{center}
    {\LARGE\bf Neural Simulation-Based Inference for space-time Gaussian processes}
\end{center}
  \medskip
} \fi

\bigskip
\begin{abstract}
In spatio-temporal data analysis, estimating covariance parameters for Gaussian random fields via maximum likelihood estimation (MLE) is computationally prohibitive for large datasets. Following recent advances in simulation-based inference techniques, we propose two neural network architectures — Neural Bayes Estimation (NBE) and Neural Ratio Estimation (NRE) — to learn the parameters of Gaussian spatio-temporal Gneiting-Matérn models with advection. The networks are trained on simulations generated directly from the underlying Gaussian random field model. A thorough simulation study demonstrates strong performance across one-dimensional time and both one-dimensional space and two-dimensional space, yielding high accuracy and reliable uncertainty quantification for both NBE and NRE. Through extensive offline training, the framework achieves fully amortized inference, replacing costly optimization with a single forward pass. Finally, we apply our framework to solar radiation prediction and short-term forecasting. By adopting a sliding-window approach, the model accounts for temporal non-stationarity while retaining this key advantage: achieving near-instantaneous parameter inference and forecasting.
\end{abstract}

\noindent%
\vfill

\newpage
\spacingset{1.8} 

\section{Introduction}

Geostatistics aims at modeling natural phenomena that evolve in space and time and predicting
their behavior at unobserved locations from observations that are distributed irregularly. 
It offers numerous  applications whether in the study of air pollution \citep{Sun2019, Idir2021}, solar radiation \citep{clarotto2024spde} or rainfalls \citep{hess-30-4799-2026}. 
Most geostatistical models rely on Gaussian processes \citep{cressie2011statistics}, as they allow for an easy prediction of the variable of interest at unobserved locations together with the computation of the prediction variance. In this context,
the observations are assumed to be derived from a particular realization of a Gaussian process. The second order moment, that shapes the regularity of the process,
is typically chosen among parametric families of covariance functions. The parameters of the covariance functions are classically estimated by maximizing the likelihood of the data \citep{mardia1984maximum}. In practice, however, Maximum Likelihood Estimation (MLE) becomes a computational bottleneck as the amount of data increases \citep{banerjee2003hierarchical} — an issue that frequently arises in a spatio-temporal context.

A large number of methods attempt to accelerate the computation of the likelihood, as reviewed in \cite{heatonCaseStudyCompetition2019}.
Low rank approximations \citep{Cressie2022} and predictive processes \citep{banerjee2008gaussian} simplifies likelihood computations by reducing them to the inversion of covariance matrices of much smaller size. Such decompositions form the basis of several extensions, particularly in the spatial setting \citep[e.g][]{Finley2009, Cressie2022, nguyen2017multivariate, Huang2018, ZammitMangion2021} but also in the spatio-temporal setting \citep[e.g][]{wikle2019spatio, ZammitMangion2021}. 

It is also possible to exploit such decompositions through precision matrices \citep{lindgren2011explicit, nychka2015multiresolution, Cressie2022}, i.e. the inverses of covariance matrices. Following mathematical and algorithmic advances in the spatial framework \citep{fuglstad2015does, pereira2019efficient, pereira2022geostatistics}, \citet{clarotto2024spde} and \citet{pereira2026prediction} extended these methods to spatio-temporal settings by incorporating advection and diffusion terms into the underlying partial differential equation. Sparse precision matrices can also be obtained through alternative constructions. A classical approach is that of \citet{vecchia1988estimation}, which approximates the joint distribution of spatial random vectors by conditioning each location on a small subset of its nearest previously observed neighbors \citep{datta2016hierarchical,katzfuss2021general}.

Most of these methods may be considered as fast approximations over the MLE but might still suffer from a prohibitive computation time as the amount of data increases. This limitation has motivated the development of likelihood-free inference approaches, which avoid explicit evaluation of the likelihood function. These methods rely instead on the ability to efficiently simulate data from the model for a given set of parameters. A prominent example is approximate Bayesian computation (ABC) \citep{sisson2018handbook}, which performs inference by comparing observed data with simulated realizations and retaining parameter values that yield sufficiently small discrepancies under a chosen distance metric.

More generally, the simulation-based perspective has led to a broader class of methods referred to as simulation-based inference (SBI) \citep{Cranmer2020, Deistler2025}. Recent advances in SBI increasingly leverage deep neural networks, to learn mappings between observations and parameters or to approximate posterior distributions directly \citep{Cranmer2020, Deistler2025}. This framework relies on constructing a training dataset of parameters–simulation pairs, where the parameters are sampled from a given prior distribution and synthetic observations are generated from the model. A neural network is then trained on this dataset to learn either the inverse mapping from data to parameters or an approximation of the posterior distribution. Once trained, inference becomes highly efficient, since it only requires forward passes through the network. This leads to an \emph{amortized} effect \citep{ZammitMangion2025}, where the expensive part is restricted to the phase of simulation and training, while the cost of inference is essentially independent of the number of observations.

The use of SBI in geostatistics has recently started to gain prominence due to its effective applications. The range of applications has been reviewed in \cite{ZammitMangion2025} for the spatial case. Among them, the works of \cite{SainsburyDale2025} and \cite{walchessen2024neural} have provided useful inspiration for this work. To our knowledge, SBI has never been used in a spatio-temporal framework.  

Following these recent studies, we propose two neural SBI methods designed to estimate the parameters of spatio-temporal Gaussian models characterized by non-separability and advection. The networks are trained on multiple simulations generated directly from the underlying Gaussian random field model. We employ  convolutional neural network (CNN) architectures, which extract spatio-temporal information from scattered observations by mapping them onto a discrete grid. We evaluate and compare the performance of these different architectures in terms of estimation accuracy, uncertainty quantification, and computational cost.

The structure of the paper is as follows: in Section~\ref{section:background}, we introduce the general framework of SBI, together with two specific approaches: Neural Bayes Estimators and Neural Ratio Estimators. We then set the theoretical context of geostatistics, including spatio-temporal modeling, simulation, and prediction. Section~\ref{section:contribution} presents our contribution, in particular the choice of the spatio-temporal model and the method used to simulate the training data. 
We then describe how to embed spatio-temporal data into a summary representation using a CNN. Section~\ref{section:sim_studies} outlines the experimental design and presents results within a fully simulated framework. In Section~\ref{section:applications}, we show the efficiency of our method through an application to real-world observations of solar radiation for photovoltaic power forecasting.
Finally, Section~\ref{section:conclusion} concludes the paper with a discussion of limitations and directions for future work.

\section{Background}\label{section:background}

\subsection{Simulation-Based Inference}

Let $\boldsymbol{y}_\text{obs} = \left( y_1, ..., y_n \right) \in \mathcal{Y} \subset \mathbb{R}^{n}$ be an observation vector assumed to be sampled from a realization of a model parametrized by an unknown parameter vector $\boldsymbol{\theta}_\text{obs}= \left( \theta_1, ..., \theta_d \right) \in \Theta \subset \mathbb{R}^d$.
Simulation-based inference (SBI) provides a general framework for parameter estimation by leveraging repeated simulations of the model rather than explicit likelihood computations \citep{Cranmer2020, ZammitMangion2025, Deistler2025}. The key assumption underlying these methods is that simulation from a given model parameterized by a vector $\boldsymbol{\theta}$ is reasonably fast, even when direct likelihood evaluation is not. 

\subsubsection{General workflow}


Fast simulations are then exploited in a simulation-based strategy: the model is sampled $N > 0$ times to construct a dataset of simulated observations paired with their associated parameter values. This dataset is generated as follows:

\begin{itemize}
    \item $\boldsymbol{\theta}^{(1)}, \dots, \boldsymbol{\theta}^{(N)} \in \Theta$ are drawn from chosen prior distribution $p(\boldsymbol{\theta})$,
    \item $\boldsymbol{y}^{(1)}, \dots, \boldsymbol{y}^{(N)} \in \mathcal{Y}$ are simulated from the chosen model, parametrized by $\boldsymbol{\theta}^{(1)}, \dots, \boldsymbol{\theta}^{(N)}$ respectively. 
\end{itemize}

The goal of SBI is to leverage a simulation-only training dataset to train a neural network that approximates the relationship between observations and model parameters. In this setting, the network becomes capable of mapping a given observation (whether simulated or real) to one or a posterior distribution of corresponding parameter values. Although this training phase is computationally intensive, it is performed offline. Once the network has been trained, parameter estimation for new observations $\boldsymbol{y}_\text{obs}$ is extremely fast. Depending on the chosen method (cf. Section~\ref{subsubsection:NBE} or \ref{subsubsection:NRE}), inference may require only a single forward pass through the neural network.

These methods are therefore referred to as \emph{amortized}, since the cost of simulation and training is incurred only once, after which inference on new observations can be performed at very low computational cost. The workflow can be taken a step further using sequential SBI \cite{Papamakarios2016, lueckmann2017flexible, greenberg2019automatic}, in which the simulation-training process is repeated several times and the prior is updated based on the estimated posterior from the previous iteration. This technique improves accuracy but completely loses the amortized nature.

\subsubsection{Method 1 : Neural Bayes Estimators (NBE)}\label{subsubsection:NBE}

For the use case of point estimation, a common method is the Neural Bayes Estimator (NBE). In this framework, the neural network takes the data as input and directly outputs estimates of the model parameters (see e.g. \cite{flagel2019unreasonable, gerber2021fast, rudi2022parameter, lenzi2023neural, sainsbury2024likelihood, SainsburyDale2025}). To achieve this, a summary representation of the data is mapped to the parameter space using a neural network trained with a loss such as the mean squared error (MSE) or the mean absolute error (MAE). 

In the remainder of this section, we focus on the methodological aspects of NBE as well as techniques for constructing associated confidence intervals.

\paragraph{\textbf{Estimation}}

We denote the estimator as $\widehat{\boldsymbol\theta} : \mathcal{Y} \rightarrow \Theta$ which is a mapping from the observation space $\mathcal{Y}$  to the parameter space $\Theta$. A loss function $L : \Theta \times \Theta \rightarrow \mathbb R_+$, $L(\boldsymbol\theta, \widehat{\boldsymbol{\theta}}(\boldsymbol{y}))$ evaluates an estimator $\widehat{\boldsymbol{\theta}}(\cdot)$ for a parameter $\boldsymbol{\theta}$ and a realization $\boldsymbol{y}$. 
For a given $\boldsymbol{y}$, by Bayes rule, the expectation of the posterior loss is 
\begin{equation}
\int_\Theta L(\boldsymbol\theta, \widehat{\boldsymbol{\theta}}(\boldsymbol{y})) p(\boldsymbol{\theta} | \boldsymbol{y}) \diff \boldsymbol{\theta} = \int_\Theta L(\boldsymbol\theta, \widehat{\boldsymbol{\theta}}(\boldsymbol{y})) p(\boldsymbol{y}| \boldsymbol{\theta}) \frac{p(\boldsymbol{\theta})}{p(\boldsymbol{y})} \diff \boldsymbol{\theta},
\end{equation} 
where $p(\boldsymbol{y}| \boldsymbol{\theta})$ is the likelihood function, $p(\boldsymbol{y})$ is the evidence and $p(\boldsymbol{\theta})$ is a chosen prior distribution.
The Bayes estimator $\widehat{\boldsymbol{\theta}}^*(\cdot)$ minimizes this risk 
\begin{equation}
\widehat{\boldsymbol{\theta}}^*(\cdot) = \argmin_{\widehat{\boldsymbol{\theta}}(\cdot)} \int_{\Theta} \int_{\mathcal{Y}} L(\boldsymbol{\theta}, \widehat{\boldsymbol{\theta}}(\boldsymbol{y})) p(\boldsymbol{y}| \boldsymbol{\theta}) p(\boldsymbol{\theta}) \diff \boldsymbol{y} \diff \boldsymbol{\theta}.
\end{equation}
Let $\widehat{\boldsymbol{\theta}}_{\boldsymbol{\delta}}(\cdot)$ be a neural network with weights $\boldsymbol\delta$. Assuming the network is flexible enough, the Bayes estimator can be approximated by the \textit{neural Bayes estimator} $\widehat{\boldsymbol{\theta}}_{\boldsymbol{\delta}^*}(\cdot)$ where
\begin{equation}
\boldsymbol{\delta}^* = \argmin_{\boldsymbol{\delta}} \sum_{l=1}^N L(\boldsymbol\theta^{(l)}, \widehat{\boldsymbol{\theta}}_{\boldsymbol{\delta}}(\boldsymbol{y}^{(l)})),
\end{equation} 
given a set of simulations $(\boldsymbol\theta^{(l)},\boldsymbol{y}^{(l)})$, $l \in \{1, ..., N\}$. This minimization is classically carried out by stochastic gradient descent where the gradients are computed by back-propagation. After training, estimating $\boldsymbol{\theta}_{\text{obs}}$ from $\boldsymbol{y}_{\text{obs}}$ only requires a forward pass in the neural network.

\paragraph{Uncertainty quantification}


In the NBE setting, two main approaches have been explored to address uncertainty quantification for model parameters. A first class of methods relies on bootstrap-based strategies, as in \cite{sainsbury2024likelihood, ZammitMangion2025}. In these works, the \emph{deepset} architecture—originally designed to handle replicated datasets—is leveraged to perform non-parametric bootstrap inference. 

An alternative approach is parametric bootstrapping \citep{efron2012bayesian, Cranmer2020}. Starting from a point estimate $\widehat{\boldsymbol{\theta}}$, $B$ simulated datasets are generated from the fitted model. Each simulated dataset is then passed through the trained network to obtain $B$ corresponding parameter estimates. Empirical confidence intervals are finally constructed using the resulting quantiles. In our setting, this approach is appealing due to the fast simulation capability of the underlying model. However, it partially breaks the amortized nature of SBI, since additional simulations are required for each new observation.

Another class of methods is based on quantile losses, inspired by \cite{SainsburyDale2025} and \cite{padilla2022quantile}. The starting point is to interpret the neural network estimator $\widehat{\boldsymbol{\theta}}(\cdot)$ as targeting the marginal posterior median by default. To achieve this, the network is trained using the MAE loss
\[
L_M(\boldsymbol{\theta}, \widehat{\boldsymbol{\theta}}) = \lVert \boldsymbol{\theta} - \widehat{\boldsymbol{\theta}} \rVert_1,
\]
which targets the conditional median.

A set of quantile levels $\{\alpha_1, \dots, \alpha_T\}$ is selected below the median, with $0.5 > \alpha_1 > \cdots > \alpha_T$, and a set $\{\beta_1, \dots, \beta_{T'}\}$ above the median, with $0.5 < \beta_1 < \cdots < \beta_{T'}$. The aim is then to estimate all the quantile levels. Each quantile is learned by extending the base network $\widehat{\boldsymbol{\theta}}(\cdot)$ as follows:
\begin{align*}
    \widehat{\boldsymbol{\theta}}^{(\alpha_t)}(\cdot) = \widehat{\boldsymbol{\theta}}(\cdot) - \sum_{i=1}^{t} {g}(\nu^{(\alpha_i)}(\cdot)), \quad t \in \{1, ..., T\} \\ \widehat{\boldsymbol{\theta}}^{(\beta_t)}(\cdot) = \widehat{\boldsymbol{\theta}}(\cdot) + \sum_{i=1}^{t} {g}(\nu^{(\beta_i)}(\cdot)), \quad t \in \{1, ..., T'\}
\end{align*}
    
where $\nu^{(\cdot)}(\cdot)$ are (smaller) variant of the main network and $g(\cdot)$ is a non-negative function. 
This approach yields valid quantile estimates as
$$\widehat{\boldsymbol{\theta}}^{(\alpha_T)} < ...< \widehat{\boldsymbol{\theta}}^{(\alpha_1)} < \widehat{\boldsymbol{\theta}} < \widehat{\boldsymbol{\theta}}^{(\beta_1)} < ... <\widehat{\boldsymbol{\theta}}^{(\beta_{T'})}.$$
All the networks, including $\widehat{\boldsymbol{\theta}}(\cdot)$, are trained simultaneously on the loss 
\begin{equation}
    L(\boldsymbol{\theta},\widehat{\boldsymbol{\theta}}, \widehat{\boldsymbol{\theta}}^{(\alpha_1)}, ..., \widehat{\boldsymbol{\theta}}^{(\alpha_T)}, \widehat{\boldsymbol{\theta}}^{(\beta_1)}, ... \widehat{\boldsymbol{\theta}}^{(\beta_T')}) = L_M(\boldsymbol{\theta},\widehat{\boldsymbol{\theta}}) + \sum_{t=1}^{T} L_{\alpha_t}(\boldsymbol{\theta},\widehat{\boldsymbol{\theta}}^{(\alpha_t)}) + \sum_{t=1}^{T'} L_{\beta_t}(\boldsymbol{\theta},\widehat{\boldsymbol{\theta}}^{(\beta_t)}),
\end{equation}
where $L_M(\cdot,\cdot)$ is the MAE and $L_{\alpha_t}(\cdot,\cdot)$, $t \in \{1,...,T\}$ (resp. $L_{\beta_t}(\cdot,\cdot)$, $t \in \{1,...,T'\}$) are the quantile losses 
\begin{equation}
    L_{\alpha_t}(\boldsymbol{\theta},\widehat{\boldsymbol{\theta}}^{(\alpha_t)}) = (\widehat{\boldsymbol{\theta}}^{(\alpha_t)} - \boldsymbol{\theta})(\mathbb{I}_{\widehat{\boldsymbol{\theta}}^{(\alpha_t)} > \boldsymbol{\theta}} - \alpha_t),
\end{equation}
where $\mathbb{I}(\cdot)$ denotes the indicator function.


The quantile-loss approach yields a fully amortized inference framework. In contrast to bootstrap-based methods, which require additional simulations at inference time, this strategy shifts the computational burden to the training phase, leading to a fully amortized workflow. The choice between these two approaches therefore depends on the specific application context, in particular the trade-off between offline training cost and online computational efficiency.

\subsubsection{Method 2 : Neural Ratio Estimators (NRE)}\label{subsubsection:NRE}

In certain settings within SBI, it is also possible to directly approximate the likelihood with a Neural Likelihood Estimator (NLE). This can be advantageous when the posterior is highly concentrated and simulations are computationally expensive \cite{radev2023jana, papamakarios2019sequential, Deistler2025}. 

However, in most practical applications, rather than the likelihood itself, the likelihood ratio is estimated (e.g. \cite{thomas2022likelihood, delaunoy2022towards, walchessen2024neural}). One common way of approximating this ratio is to formulate the problem as a binary classification task. 
Specifically, the network is trained to distinguish between pairs where the parameter and the data are consistent with one another—meaning the parameter could have generated the observed data—and pairs where the parameter–data combination is mismatched, i.e., the parameter is unrelated to the data. The resulting binary cross-entropy (BCE) loss yields an estimate of the likelihood ratio, as detailed below.

Extensions to multi-class classification have also been proposed to estimate likelihood-to-evidence ratios, for example in \cite{durkan2020contrastive} and \cite{miller2022contrastive}. Multi-class classification is obtained by training a neural network to identify the true pair parameter-data among a set of $K$ alternative contrastive candidates. While being more expressive, these approaches typically introduce additional training hyperparameters, inducing reduced robustness across different problem settings. This is particularly problematic in spatial and spatio-temporal contexts, where maintaining a fully amortized and widely applicable framework is desirable.

The binary classification formulation is illustrated in Figure~\ref{fig:architecture_nre} and is detailed in the remainder of this section.

\paragraph{Estimation}

For any parameter $\boldsymbol{\theta} \in \Theta$ and realization $\boldsymbol{y} \in \mathcal{Y}$, the likelihood-to-evidence ratio is defined as in \cite{walchessen2024neural}:
\begin{equation}
r(\boldsymbol{y}, \boldsymbol{\theta}) =
\frac{\mathcal{L}(\boldsymbol{\theta} \mid \boldsymbol{y})}{p(\boldsymbol{y})},
\end{equation}
where, $\mathcal{L}(\boldsymbol{\theta} \mid \boldsymbol{y})$ denotes the likelihood function and $p(\boldsymbol{y})$ the model evidence.

Following \cite{Hermans2020, thomas2022likelihood, walchessen2024neural}, this ratio is learned via a binary classification problem. A neural classifier $h : \mathcal{Y} \times \Theta \rightarrow [0,1]$ is trained to distinguish between two types of pairs. In the first case ($D = 1$), the pair $(\boldsymbol{y}, \boldsymbol{\theta})$ is jointly distributed according to the model : $p(\boldsymbol{y}, \boldsymbol{\theta} \mid D=1) = p(\boldsymbol{y} \mid \boldsymbol{\theta}) p(\boldsymbol{\theta})$.
In the second case, the parameters and data are sampled independently: $p(\boldsymbol{y}, \boldsymbol{\theta} \mid D=0) = p(\boldsymbol{y}) p(\boldsymbol{\theta})$.

Under this formulation, the optimal classifier $h^*(\cdot,\cdot)$ is obtained by minimizing the Bayes risk, as shown in \cite{ZammitMangion2025}:
\begin{equation}
\begin{split}
    h^*(\cdot,\cdot) 
    &= \argmin_{h(\cdot, \cdot)} \sum_{d \in \{0,1\}} P(D=d) \int_{\Theta} \int_{\mathcal{Y}} L_{\text{ce}}(d, h(\yth)) P((\yth) |D=d) \diff \boldsymbol{y} \diff \boldsymbol{\theta},\\
\end{split}
\end{equation}
where $L_{\text{ce}}(d,\widehat d) = -d\log(\widehat d) - (1-d)\log(1-\widehat d)$ is the BCE.

On the other hand, by Bayes' Rule \citep{walchessen2024neural}, for any given $\boldsymbol{\theta}$ and $\boldsymbol{y}$, we obtain 
\begin{equation*}
    h^*(\yth) 
    = P(D=1|(\yth)) 
    = \frac{r(\yth)}{r(\yth)+1}.
\end{equation*}
The second equality holds if we assume that the two classes are balanced, i.e. $P(D=1) = P(D=0) = \frac{1}{2}$. This can be achieved by concatenating the dataset $(\boldsymbol\theta^{(l)},\boldsymbol{y}^{(l)})$, $l \in \{1, ..., N\}$ with the dataset $(\boldsymbol\theta^{(\pi(l))},\boldsymbol{y}^{(l)})$, $l \in \{1, ..., N\}$ where $\pi(\cdot)$ is a random permutation. For every $\boldsymbol{\theta} \in \Theta$ and $\boldsymbol{y} \in \mathcal{Y}$, an approximation of the likelihood-to-evidence ratio is
\begin{equation}
    r(\yth) = \frac{h^*(\yth)}{1-h^*(\yth)}.
\end{equation}

The optimal classifier $h^*$ is unknown but can be approximated by $h_{\boldsymbol{\delta}}(\cdot)$, a neural network with weights $\boldsymbol\delta$. The \textit{Neural ratio estimator} is $h_{\boldsymbol{\delta}^*}(\cdot)$ with 
\begin{equation} \label{eq:delta}
\boldsymbol{\delta}^* = \argmin_{\boldsymbol{\delta}} \sum_{l=1}^N L_{\text{ce}}(1, h_{\boldsymbol{\delta}}(\boldsymbol\theta^{(l)}, \boldsymbol{y}^{(l)})) + \sum_{l=1}^N L_{\text{ce}}(0, h_{\boldsymbol{\delta}}(\boldsymbol\theta^{(\pi(l))}, \boldsymbol{y}^{(l)})).
\end{equation}

Inferring $\boldsymbol{\theta}_{\text{obs}}$ from $\boldsymbol{y}_{\text{obs}}$ can be formulated as the maximization of the estimated likelihood-to-evidence ratio $r_{\boldsymbol{\delta}^*}(\cdot, \cdot)$:
\begin{equation}
    \widehat{\boldsymbol{\theta}}_{\text{obs}} = \argmax_{\boldsymbol{\theta}} r_{\boldsymbol{\delta}^*}(\boldsymbol{\theta}, \boldsymbol{y}_{\text{obs}}).
\end{equation}


Starting from an initial guess $\boldsymbol{\theta}_0$, iterative updates are performed using optimization algorithms. Gradient-based techniques, such as gradient descent \citep{ruder2016overview}, are appealing since the neural network is fully differentiable with respect to its inputs. Although this optimization approach is computationally efficient, it only provides a point estimate and does not directly yield measures of uncertainty, such as confidence intervals.

\paragraph{Uncertainty quantification}

In order to perform uncertainty quantification (UQ), \cite{walchessen2024neural} construct a regular grid over the parameter space $\Theta$ and evaluate the neural network at each grid point $\boldsymbol{\theta}$. This yields a discretized likelihood surface over $\Theta$. The point estimate $\widehat{\boldsymbol{\theta}}_{\text{obs}}$ is then obtained by maximizing this surface. Based on this estimator and the full evaluated grid, the authors construct a statistical test by inverting the likelihood-ratio test (see \cite{walchessen2024neural, Casella2024}, Section 10.4), which allows the derivation of a confidence region.

However, unlike the UQ approaches discussed in the NBE framework (Section~\ref{subsubsection:NBE}), this confidence region does not directly provide marginal confidence intervals for each component of $\widehat{\boldsymbol{\theta}}_{\text{obs}}$. An alternative strategy is to rely on Markov chain Monte Carlo (MCMC) methods, as in \cite{Hermans2020}, which yield an approximation of the full posterior distribution. From this posterior approximation, point estimates such as the maximum a posteriori (MAP) estimator can also be derived.

This MCMC-based approach is not without limitations. First, it is sensitive to the choice of sampling algorithm and associated hyperparameters. While the No-U-Turn Sampler (NUTS) \citep{hoffman2014no} is often considered state-of-the-art due to its limited number of tuning parameters, it still requires careful implementation. Second, the computational cost remains substantial, as running MCMC and computing MAP estimates for each new observation can be prohibitive.

More generally, the choice of the inference strategy—within NRE, but also across NBE and other SBI approaches—depends strongly on the application context. In particular, it reflects a trade-off between fully amortized inference and per-dataset computational cost, which is itself driven by the cost of simulations and the number of observational datasets to be analyzed. A detailed comparison of posterior estimation strategies (NPE, NLE, and NRE) is provided in \cite{Deistler2025}. Within a spatio-temporal framework, presented in the following section, such a trade-off remains an open question.

\subsection{Geostatistical framework}

The following section describes our spatio-temporal setting, meaning the Gaussian random field modeling and the traditional methods used for inference and prediction. We also present an approach to perform fast simulations of Gaussian random fields.. 

\subsubsection{Spatio-temporal Gaussian model}

Let $y(\boldsymbol{s}_i, t_j) : \cal D \times \cal T \rightarrow \mathbb R $ be noisy observations at spatial points $(\boldsymbol{s}_1, ..., \boldsymbol{s}_n)$ of $\mathcal{D} \subset \mathbb{R}^d$, $d > 0$ and time steps $(t_1, ..., t_m)$ of $\mathcal T \subset \mathbb R$. 
Without loss of generality, we assume a fixed spatial sampling over time, i.e. the same spatial locations are observed at each time step.
To model these data, for each $i \in \{1, ..., n\}$ and $j \in \{1, ..., m\}$, we suppose that the observation $y(\boldsymbol{s}_i, t_j)$ is a realization of the random variable 
\begin{equation} 
    Y(\boldsymbol{s}_i, t_j) = \mu(\boldsymbol{s}_i, t_j) + Z(\boldsymbol{s}_i, t_j) + \tau\epsilon_{i,j}, \quad \tau > 0,
    \label{eq:model_ST}
\end{equation}
where $\mu$ is the deterministic trend, 
$\epsilon_{i,j} \sim \mathcal N (0,1)$ is an iid Gaussian noise, and
$Z$ is a zero-mean Gaussian random field with stationary covariance function $C:\cal D \times \cal T \rightarrow \mathbb R$,
\begin{equation}C(\boldsymbol{h},u) = \text{Cov}\left(Y(\boldsymbol{s} + \boldsymbol{h}, t + u), Y(\boldsymbol{s}, t)\right), \quad \forall (\boldsymbol{s}, t) \in (\mathcal{D} \times \mathcal{T}).\end{equation}
The covariance function dictates how two points in space and time influence each other. We suppose stationarity, therefore the covariance only depends on spatial and temporal distances $\boldsymbol{h}$ and $u$, not on space-time points themselves.

Specifying a covariance function implies the existence of the covariance matrix $\boldsymbol{\Sigma}$ which assigns $C(\cdot, \cdot)$ to each pair of observed points. To define it, we introduce the matrix notation
\begin{equation*} \boldsymbol{Z} = \begin{pmatrix}
    \boldsymbol{Z}_{t_1} \\ \boldsymbol{Z}_{t_2} \\ \cdots \\ \boldsymbol{Z}_{t_m}
\end{pmatrix} \quad \text{ with} \quad 
\boldsymbol{Z}_{t_j} =  \begin{pmatrix}
    {Z}(\boldsymbol{s}_1, t_j) \\ {Z}(\boldsymbol{s}_2, t_j) \\ \cdots \\ {Z}(\boldsymbol{s}_n, t_j)
\end{pmatrix}, \quad \forall j \in \{1, ..., m\}.
\end{equation*} 
Same goes for $\boldsymbol{\mu}$ an $\boldsymbol{Y}$. Then, we have $\boldsymbol{Z} \sim \mathcal{N}(\boldsymbol{0}, \boldsymbol{\Sigma})$ and $\boldsymbol{Y} \sim \mathcal{N}(\boldsymbol{\mu}, \boldsymbol{\Sigma} + \tau^2 \boldsymbol{I})$ where
\begin{equation*}
\boldsymbol{\Sigma} = \begin{pmatrix}
    \ddots & \cdots & \iddots \\
    \cdots & \text{Cov}(Z(\boldsymbol{s}_i,t_j),Z(\boldsymbol{s}_{i'},t_{j'})) & \cdots \\
    \iddots & \cdots & \ddots
\end{pmatrix}_{\substack{\{1 \le i \le n, 1 \le j \le m\} \\ \{1 \le i' \le n, 1 \le j' \le m\}}}.
\end{equation*}
In practice, we choose a covariance function $C_{\boldsymbol{\theta}}(\cdot, \cdot)$ parameterized by $\boldsymbol{\theta} \in \Theta \subset \mathbb{R}^p$, where $\Theta$ is the parameter space and $p$ is the number of parameters.
We denote the spatio-temporal observation vector $\boldsymbol{y}_\text{obs} = \left( y(\boldsymbol{s}_1, t_1), ..., y(\boldsymbol{s}_n, t_m) \right) \in \mathcal{Y} \subset \mathbb{R}^{n\times m}$.
 The goal is to estimate the parameters $\boldsymbol{\theta}_\text{obs}$ of the covariance function that generated the observations $\boldsymbol{y}_\text{obs}$.

\subsubsection{Standard inference method}

The classical approach to estimate the parameters $\boldsymbol{\theta}_\text{obs}$ from $\boldsymbol{y}_\text{obs}$ is Maximum Likelihood estimation (MLE). The evaluation of the log-likelihood function
\begin{equation}
\begin{split}
    \mathcal{L}(\boldsymbol{\theta} | \boldsymbol{y_\text{obs}}) = -\frac{nm}{2}\log(2\pi) & - \frac{1}{2}\log\det(\boldsymbol{\Sigma}_{\boldsymbol{\theta}}) - \frac{1}{2} \|\boldsymbol{y_\text{obs}} -  \boldsymbol{\mu}\|^2_{\boldsymbol{\Sigma^{-1}_{\boldsymbol{\theta}}}} \\
    \text{with } \|v\|_\mathbf{A} := v^\top \mathbf{A}^{-1}v
\end{split}
\end{equation}
requires solving systems involving $\boldsymbol{\Sigma}_{\boldsymbol{\theta}}$, the covariance matrix induced by $C_{\boldsymbol{\theta}}(\cdot, \cdot)$ over all pairs of observed locations. In practice, this computation becomes prohibitive as the number of points—and consequently the size of the matrix—increases. Moreover, in order to maximize the log-likelihood, these operations must be repeated at each iteration of the optimization algorithm.


\subsubsection{Spatio-temporal prediction}

Let $\widehat{\boldsymbol{\theta}}_\text{obs}$ be an estimation of the parameters behind the observations $\boldsymbol{y}_{\text{obs}}$ at locations $(\boldsymbol{s}_{i},t_{j}), i \in \{1, ..., n\}, j \in \{1, ..., m\}$. 
Kriging is a classic tool in geostatistics to perform prediction from the estimated model at unmeasured locations $(\boldsymbol{s'}_{i'},t'_{j'}), i' \in \{1, ..., n'\}, j' \in \{1, ..., m'\}$ \citep{chiles2012geostatistics}. 
We first define $\boldsymbol{Y}'$ the vector of the random variable at unobserved locations. We also denote $\boldsymbol{\mu}_\text{obs}$ and $\boldsymbol{\mu}_\text{pred}$ the estimated trends of the observation and prediction points respectively. The covariance matrix $\boldsymbol{\Sigma}_{\text{obs-obs}}$ contains the values of $C_{\widehat{\boldsymbol{\theta}}}(\cdot,\cdot)$ evaluated between all pairs of observation locations, while $\boldsymbol{\Sigma}_{\text{pred-obs}}$ contains the covariances between prediction locations and observation locations.
The prediction by kriging $\boldsymbol{y}_{\text{pred}}$ is the conditional expectation of $\boldsymbol{Y}'$ given the observations, i.e.
\begin{equation}
\begin{split}
        \boldsymbol{y}_{\text{pred}} =  \mathbb{E}[\boldsymbol{Y}' | \boldsymbol{Y} = \boldsymbol{y}_{\text{obs}}] = \boldsymbol{\mu}_\text{pred} + \boldsymbol{\Sigma}_\text{pred-obs} & \boldsymbol{V}^{-1}_\text{obs-obs}  (\boldsymbol{y}_{\text{obs}}-\boldsymbol{\mu}_\text{obs}) \\  \quad \text{with } \boldsymbol{V}_\text{obs-obs} = \boldsymbol{\Sigma}_\text{obs-obs} + \tau^2\boldsymbol{I}, \qquad \tau > 0.
\end{split}
\end{equation}
This conditional expectation is linear and may therefore produce results that are less reliable in regions with sparse or missing observations. A measure of uncertainty on the prediction is given by the kriging variance
\begin{equation}
    \mathbb{V}[\boldsymbol{Y}' | \boldsymbol{Y} = \boldsymbol{y}_{\text{obs}}]= \boldsymbol{\Sigma}_\text{pred-pred} -\boldsymbol{\Sigma}_\text{pred-obs}\boldsymbol{V}^{-1}_\text{obs-obs}\boldsymbol{\Sigma}_\text{obs-pred} .
\end{equation}

\subsubsection{Simulation} \label{subsection:spectral_sim}

Simulating a GRF provides a representation of the variability across different realizations. The standard approach involves a Cholesky decomposition of the covariance matrix $C_{\boldsymbol{\theta}}(\cdot,\cdot)$ which has computational cost $O(n^3)$ \citep{allard2025modeling}, and limits widely the number of points to be simulated. As for likelihood evaluation, low-rank \citep{romary2013incomplete} and sparse approximation methods \citep{jurek2022hierarchical} have been developed as alternatives to enable faster simulation.

An effective and computationally scalable approach is the spectral method, which decomposes a GRF into a superposition of spectral waves \citep{shinozuka1971simulation, matheron1973intrinsic, lantuejoul2001geostatistical}. 
More recently, \cite{allard2020simulating, allard2022fully} extended the spectral approach to the spatio-temporal setting, in particular for the Gneiting class of covariance functions \citep{gneiting2002nonseparable}, further characterized in \citep{zastavnyi2011characterization}. Grounded in Bochner's theorem \citep{bochner1955}, which states that a valid continuous covariance function $C$ is the Fourier transform of a nonnegative spectral measure $F$,
\begin{equation}
C(\boldsymbol{h}, u) = \int_{\mathcal{D}} \int_{\mathcal{T}} e^{i \langle \boldsymbol{\omega},\boldsymbol{h} \rangle + i \tau u} F(d \boldsymbol{\omega}, d\tau ), \quad (\boldsymbol{h}, u) \in \mathcal{D} \times \mathcal{T},
\end{equation}
the spectral method consists of sampling random frequencies according to the spectral measure and projecting the resulting waves onto the target coordinates.
The method is briefly described in Algorithm~\ref{alg:spectral_sim_conceptual}.

\begin{algorithm}[H]
\small
\caption{\small Spectral Simulation of a Spatio-Temporal Gaussian Random Field}
\label{alg:spectral_sim_conceptual}

\KwIn{Spectral distribution $F$ related to the covariance model, number of waves $L$}

\BlankLine
\For{$l = 1$ \KwTo $L$}{
    Simulate a joint space-time frequency vector $(\Omega_l, \Psi_l)$ drawn from the spectral distribution $F$\;
    Simulate a random phase $\Phi_l$ uniformly distributed on $[0, 2\pi]$\;
    Simulate an amplitude factor $U_l$ uniformly distributed on $[0, 1]$\;
}

\BlankLine
\ForEach{target space-time location $(\boldsymbol{s}, t) \in \mathcal{D} \times \mathcal{T}$}{
    Compute and return the field value as the sum of the $L$ cosine waves:
    \[
    Z(\boldsymbol{s}, t) = \sum_{l=1}^{L} \sqrt{\frac{-2 \ln(U_l)}{L}} \cos \left( \langle \Omega_l, \boldsymbol{s} \rangle + \Psi_l t + \Phi_l \right)
    \]
}
\end{algorithm}

The generation of training data by the spectral simulator is then facilitated by the simulator's speed. Moreover, this efficient simulation framework, combined with kriging, allows us to perform conditional simulations, i.e., simulations that incorporate initial observations $(\boldsymbol{s}_{i},t_{j}), i \in \{1, ..., n\}, j \in \{1, ..., m\}$ while reproducing the variability induced by the covariance function. A conditional simulation is indeed obtained as
\begin{equation}
    \boldsymbol{y}_{\text{C}} = \boldsymbol{y}_{\text{pred}} + \boldsymbol{y}_{\text{NC,sim}} - \boldsymbol{y}_{\text{NC,pred}},
\end{equation}
where $\boldsymbol{y}_{\text{pred}}$ is the kriging obtained from our observations, $\boldsymbol{y}_{\text{NC}}$ is a simulation from the estimated model at every location (both observed and unobserved) and $\boldsymbol{y}_{\text{NC,krig}}$ is the kriging obtained from the simulation at observed locations.
As $\boldsymbol{y}_{\text{pred}}$ and $(\boldsymbol{y}_{\text{NC}} - \boldsymbol{y}_{\text{NC,pred}})$ are independent, $N$ conditional simulations can be achieved at the expense of $N+1$ kriging computations and $N$ non-conditional simulations.

\subsection{Embedding spatial data in SBI}

SBI has already been applied to spatial data in previous studies. In this framework, $\boldsymbol{\theta}$ is the vector of parameters of a given spatial covariance function and $\boldsymbol{y}^{(1)}, ..., \boldsymbol{y}^{(N)}$ are simulated realizations of the spatial Gaussian random field model ${Y}(\cdot, \cdot;\boldsymbol{\theta})$.

When observations are available on a regular grid, convolutional neural networks (CNNs) are a natural choice for extracting informative summary representations. This is the case in \cite{sainsbury2024likelihood, lenzi2023neural}, where CNN-based embeddings are used for point estimation from summary vectors. In a similar setting, \cite{walchessen2024neural} also employs CNNs within an NRE framework. In these works, the embedding network is trained jointly with the network mapping the learned representation to the parameter space.

When spatial data are observed irregularly in space, alternative strategies are required. For instance, \cite{gerber2021fast} avoid neural embeddings and instead rely on empirical variograms to construct summary statistics. However, in the presence of substantial missing data, variogram estimation becomes more challenging, degrading the reliability of the resulting summaries. Another approach is proposed by \cite{SainsburyDale2025}, who use graph neural networks (GNNs) to embed irregular spatial observations. While GNNs are appealing as they naturally handle irregular sampling and capture spatial dependencies, their performance depends critically on the construction of the underlying graph. The graph should neither be too large, to avoid excessive computational costs, nor too small, to prevent loss of relevant information. 

To the best of our knowledge, SBI has not yet been investigated for spatio-temporal parameter estimation. To address this gap, we propose in the following section a method for applying SBI in a spatio-temporal framework, possibly in presence of advection or non-separability. Our approach consists of simulating spatio-temporal fields and projecting irregular spatial observations onto regular grids, allowing the use of convolutional neural networks for representation learning.

\section{Simulation-based Inference in a spatio-temporal framework} \label{section:contribution}

In this section, we propose our framework for SBI for complex spatio-temporal models. In this case, $\boldsymbol{\theta}$ is the vector of parameters of a given spatio-temporal covariance function, and we assume that it is relatively efficient to generate realizations from the corresponding model, e.g. via spectral approaches introduced in Section~\ref{subsection:spectral_sim}.


\subsection{Model} 

\label{subsection:model_matern}

We consider a zero-mean stationary Gaussian process. Let $C$ be a space-time covariance function. $C$ is said to be separable if 
\begin{equation*}
    C(\boldsymbol{h},u) = C(\boldsymbol{0},0)^{-1}C(\boldsymbol{h},0)C(\boldsymbol{0},u),
\end{equation*}
where $C(\boldsymbol{h},0)$ is purely spatial and $C(\boldsymbol{0},u)$ is purely temporal.
Among all existing space-time covariance functions, we intentionally avoid separable ones, as they structurally induce conditional independence between the spatial and the temporal fluctuations. Non-separable covariance functions are better equipped to describe complex spatio-temporal dependencies, making them much more realistic for real-world applications.

We choose $C_{ST} : \cal D \times \cal T \rightarrow \mathbb R$ as the univariate nonseparable Gneiting-Matérn covariance function \citep{allard2022fully}, derived from the Gneiting model described in \cite{gneiting2002nonseparable},
\begin{equation} 
C_{ST}(\boldsymbol{h}, u) = \sigma^2C_T\left(\frac{u}{\rho_t};\alpha, \beta\right) C_S \left( \frac{\frac{\boldsymbol{h}}{\rho_s} 
}{C_T\left(\frac{u}{\rho_t};\alpha, \beta\right)^{\gamma/d}} ; \nu \right),
\label{eq:gneiting}
\end{equation}
where $d$ is the spatial dimension.
The temporal trace is defined as
\begin{equation*}
C_{ST}(\boldsymbol{0}, u) = \sigma^2
C_T({x} ;\alpha, \beta) = \sigma^2\left(1+|x|^\alpha \right)^{-\frac{\beta}{2}d},
\end{equation*}
where $\alpha \in (0,2]$, $\beta \in (0, 1]$.
The spatial trace is the Matérn covariance function
\begin{equation*}
C_{ST}(\boldsymbol{h}, 0) = 
\sigma^2C_S(\boldsymbol{x} ; \nu) =  \sigma^2\frac{2^{1-\nu}}{\Gamma(\nu)} \left( \|\boldsymbol{x}\| \right)^\nu K_\nu \left( \|\boldsymbol{x}\| \right),
\end{equation*}
with smoothness parameter $\nu > 0$, $\Gamma(\cdot)$ the Gamma function and $K_\nu(\cdot)$ the modified Bessel function of the second kind. 
$\gamma \in [0,1]$ defines the separability, $\rho_t$ is the temporal range, $\rho_s$ is the isotropic spatial range and $\sigma^2$ is the variance. When $\gamma = 0$, the model is fully separable, but in the opposite case, i.e. $\gamma = 1$, we fall back on the original Gneiting model \citep{gneiting2002nonseparable}.

In this paper we will also consider the covariance 
\begin{equation} \label{eq:model_matern}
\tilde{C}_{ST}(\boldsymbol{h}, u) = C_{ST}(\boldsymbol{h} - \boldsymbol{v}u, u),
\end{equation}
where
$\boldsymbol{v}$ is a $d$-dimensional vector defining the advection direction, i.e. the direction in which the spatial structure shifts over time.

In the following, $\gamma$, $\nu$, $\alpha$ and $\beta$ are fixed, and $\boldsymbol{\theta}= (\boldsymbol{v}, \rho_t, \rho_s, \sigma, \tau)$ denotes the full vector of unknown parameters.
All of these parameters provide a high degree of variability, enabling the credible reproduction of complex spatio-temporal phenomena. 









\subsection{Training data generation} \label{subsection:simulation}

In order to create our training dataset, we need simulations of the spatio-temporal model \eqref{eq:model_ST} with covariance function \eqref{eq:model_matern} at spatial points $\boldsymbol{s}_i, i \in \{1, ..., n\}$  and time steps $t_j, j \in \{1, ..., m\}$.
First, simulations are performed using spectral simulation as introduced in Section~\ref{subsection:spectral_sim} \citep{allard2025modeling}. 
To obtain scattered observations, we proceed as follows. 

We choose to arrange our data into a grid rather than a graph, as the latter can be complex to set up, especially in a spatio-temporal context.
We discretize the spatial domain $\mathcal{D}$ into a regular grid $G$ consisting of $K$ disjoint cells $\{\boldsymbol{c}_1, \dots, \boldsymbol{c}_K\}$. The temporal domain is discretized on a set of regular time steps $T = (t_1, \dots, t_m)$. 
Let $y(\boldsymbol{c}_k, t_j)  : G \times T \rightarrow \mathbb R$ be a simulation of our chosen model that associate a value $y$ to each couple $(\boldsymbol{c}_k, t_j)$ for $k=1,\dots, K, j=1,\dots,m$.
To simulate irregular data in space, we define a random binary mask $\boldsymbol{\eta} = (\boldsymbol{\eta}_1, \dots, \boldsymbol{\eta}_K) \in \{0, 1\}^K$ that is shared across all time steps:
$$\tilde{y}(\boldsymbol{c}_k, t_j) = y(\boldsymbol{c}_k, t_j) \cdot \boldsymbol{\eta}_k.$$ Similarly to \cite{wang2024missing}, any missing values masked by this process are set to 0. Simulations and observations are then concatenated with their respective mask.

This choice of representation necessitates a strategy to project continuous observations onto a discrete grid. The challenge resides in the spatial dimension, which is inherently continuous. Let $\mathcal{I}_k$ be the set of indices of observations falling within the cell $\boldsymbol{c}_k$:
$$\mathcal{I}_k = \{ i \in \{1, \dots, n\} \mid \boldsymbol{s}_i \in \boldsymbol{c}_k \}.$$ 
With a fairly fine mesh size, each observation should fall into a different cell. However, as discussed in the next section, large grids lead to computational bottlenecks during both training and inference. Conversely, if a coarser mesh is chosen, multiple observations may fall within the same cell. One possible strategy to tackle this is to average the measurements within each cell, though with the risk of smoothing or breaking the underlying structure of the observations. This effect stands out particularly when the spatial range of model \eqref{eq:model_matern} is small. To preserve the original data structure without artificial smoothing, we instead chose to select a random observation from those present in the cell. Let $i_k \sim \mathcal{U}(\mathcal{I}_k)$ be the randomly selected index from the set $\mathcal{I}_k$. The final grid-based observation, denoted $\tilde{y}(\cdot, \cdot)$, is formulated as:$$\tilde{y}(\boldsymbol{c}_k, t_j) = \begin{cases} y(\boldsymbol{s}_{i_k}, t_j) & \text{if } |\mathcal{I}_k| > 0 \\ 0 & \text{if} |\mathcal{I}_k| = 0. \end{cases}$$
In the extreme case of an excessively coarse grid, there remains a risk of losing a significant amount of information from the observations. As a result, the neural network tasked with summarizing the data could end up being compromised. Therefore, a compromise must be found when choosing the grid resolution.

\subsection{Architecture of the Neural Networks}

\begin{figure}[htbp] 
    \centering
\begin{tikzpicture}[
    node distance=4cm, 
    block/.style={rectangle, draw, minimum width=3cm, minimum height=1.2cm, align=center, line width=1pt},
    arrow/.style={-{Straight Barb[length=6pt, width=8pt]}, line width=1pt}
]

  \node[block, draw=step1blue, text=black] (data) {
    $\boldsymbol{y}$
  };

  \node[block, right=of data, minimum width=2cm, draw=step1blue, text=black] (resume) {
    \textbf{ Summary}
  };

  \node[block, draw=step1blue, text=black, right=of resume, minimum width=1.5cm] (final) {
    $ \boldsymbol{\widehat{\theta}}(\boldsymbol{y})$
  };

  \draw[arrow] (data) -- (resume) 
    node[midway, above=2pt, align=center] {\textbf{Neural network}} 
    node[midway, below=2pt, align=center] {\small Convolution Layers \\ \small Max-Pooling};

  \draw[arrow] (resume) -- (final) 
    node[midway, above=2pt, align=center] {\textbf{Neural network}} 
    node[midway, below=2pt, align=center] {\small Multi Layer Perceptron};

\end{tikzpicture}
\caption{NBE neural network}
    \label{fig:architecture_nbe} 
\end{figure}
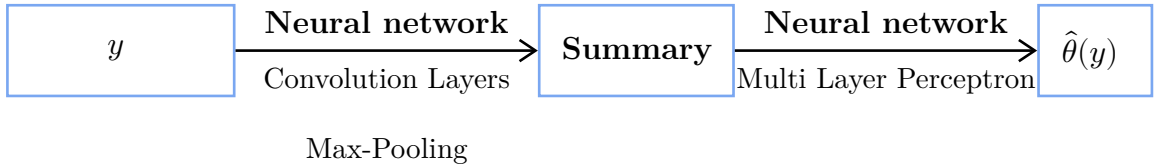

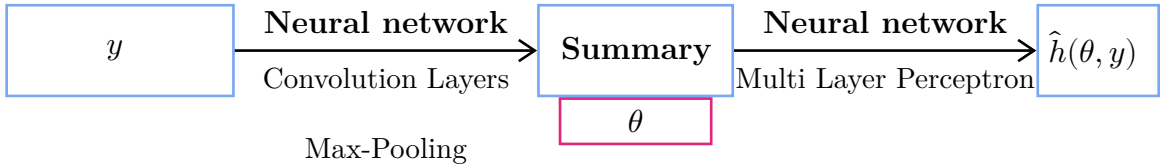
\begin{figure}[htbp] 
    \centering
\begin{tikzpicture}[
    node distance=4cm, 
    block/.style={rectangle, draw, minimum width=3cm, minimum height=1.2cm, align=center, line width=1pt},
    arrow/.style={-{Straight Barb[length=6pt, width=8pt]}, line width=1pt}
]

  \node[block, draw=step1blue, text=black] (data) {
    $\boldsymbol{y}$
  };

  \node[block, right=of data, minimum width=2cm, draw=step1blue, text=black] (resume) {
    \textbf{ Summary}
  };
  
  \node[rectangle, draw=step4pink, text=step4pink, line width=1pt, minimum width=2cm, minimum height=0.6cm, anchor=north] 
        at (resume.south) (theta) {\textcolor{black}{$\boldsymbol{\theta}$}};

  \node[block, draw=step1blue, text=black, right=of resume, minimum width=1.5cm] (final) {
    \textbf{$\widehat{h}(\boldsymbol{\theta},\boldsymbol{y})$}
  };

  \draw[arrow] (data) -- (resume) 
    node[midway, above=2pt, align=center] {\textbf{Neural network}} 
    node[midway, below=2pt, align=center] {\small Convolution Layers \\ \small Max-Pooling};

  \draw[arrow] (resume) -- (final) 
    node[midway, above=2pt, align=center] {\textbf{Neural network}} 
    node[midway, below=2pt, align=center] {\small Multi Layer Perceptron};

\end{tikzpicture}
\caption{NRE neural network}
    \label{fig:architecture_nre} 
\end{figure}

Both NBE and NRE require the input data to be transformed into an embedded representation. Given the grid-like structure of our data, CNNs are a natural architectural choice. To explicitly account for missing data locations, we include the mask as an additional input channel. Furthermore, to prevent the masked values (which are set to zero) from introducing unwanted bias, the convolutional layers are followed by a max-pooling layer that extracts the most relevant features. The output of the final max-pooling layer constitutes the summary statistics vector. The subsequent processing of this vector depends on the estimation framework: in the NBE framework the summary statistics are mapped directly into the parameter space, while in the NRE framework the summary statistics are concatenated with the parameter vector $\boldsymbol{\theta}$ to serve as inputs for the classifier. In both cases, these tasks are executed by a dense neural network that is trained jointly with the CNN. The networks' architectures are depicted in Figures~\ref{fig:architecture_nbe} and \ref{fig:architecture_nre}.

In the CNN layers, convolution is performed across all spatial and temporal dimensions simultaneously. Consequently, the temporal dimension is treated as an additional spatial dimension, yielding a joint dimensionality of $d + 1$. 
Joint spatio-temporal convolutions are essential to capture the dependencies between space and time and to detect physical phenomena, like advection, or non-separability.

However, despite being very good at capturing information, the convolutions do depend on the size of the grid. 
First, if the projected simulations' size is very large, training becomes significantly slower, as the convolution operation must be performed more frequently. Second, a large grid can lead to sparsity, meaning that the observations become extremely far apart from one another. Therefore, if the convolutional kernels are too small or the network has an insufficient number of layers, the CNN will be unable to "connect" these distant points to provide a relevant summary vector. From our experience, a grid of 50 to 100 points in each spatial dimension seems to be a good compromise for a wide range of applications.

\section{Simulation studies} \label{section:sim_studies}

In this section, we demonstrate the effectiveness of our method in a fully synthetic setting, i.e., when the neural network is trained and tested on simulations.

\subsection{Design of experiment}

\subsubsection{Parameters}


We adopt a framework with a single spatial dimension, $d = 1$, and a temporal dimension. We use the GRF model of Section~\ref{subsection:model_matern} with the Gneiting-Matérn covariance function with advection (Equation \ref{eq:model_matern}).
Our objective is to infer all the parameters of the spatio-temporal model, i.e., the advection parameter $v$, the temporal range $\rho_t$, the spatial range $\rho_s$, and the standard deviations of both the GRF $\sigma$ and the noise $\tau$. Therefore, $$\boldsymbol{\theta}= (v, \rho_t, \rho_s, \sigma, \tau).$$

\begin{longtable}[]{@{}llllll@{}}
\caption{Lower and upper bounds of each parameter used for simulations..}\label{tab:bounds}\tabularnewline
\toprule\noalign{}
& $v$ & $\rho_t$ & $\rho_s$ & $\sigma$ & $\tau$ \\
\midrule\noalign{}
\endfirsthead
\toprule\noalign{}
& $v$ & $\rho_t$ & $\rho_s$ & $\sigma$ & $\tau$ \\
\midrule\noalign{}
\endhead
\bottomrule\noalign{}
\endlastfoot
lower bound & -1 & 0.05 & 0.05 & 0.5 & 0 \\
upper bound & 1 & 0.15 & 0.3 & $\sqrt{2}$ & 0.5 \\
\end{longtable}

We begin the SBI workflow by sampling the parameters from a latin hypercube scheme with uniform priors whose bounds are shown in Table~\ref{tab:bounds}. We then perform simulations of the GRF with the library \texttt{gstlearn} \citeyear{UniMINESParisPSL2026}. As for the spatial locations, they are sampled according to the Matérn cluster process presented in the next section.

\subsubsection{Spatio-temporal sampling} \label{section:sampling}

\begin{figure}
    \centering
    \includegraphics[width=1\linewidth]{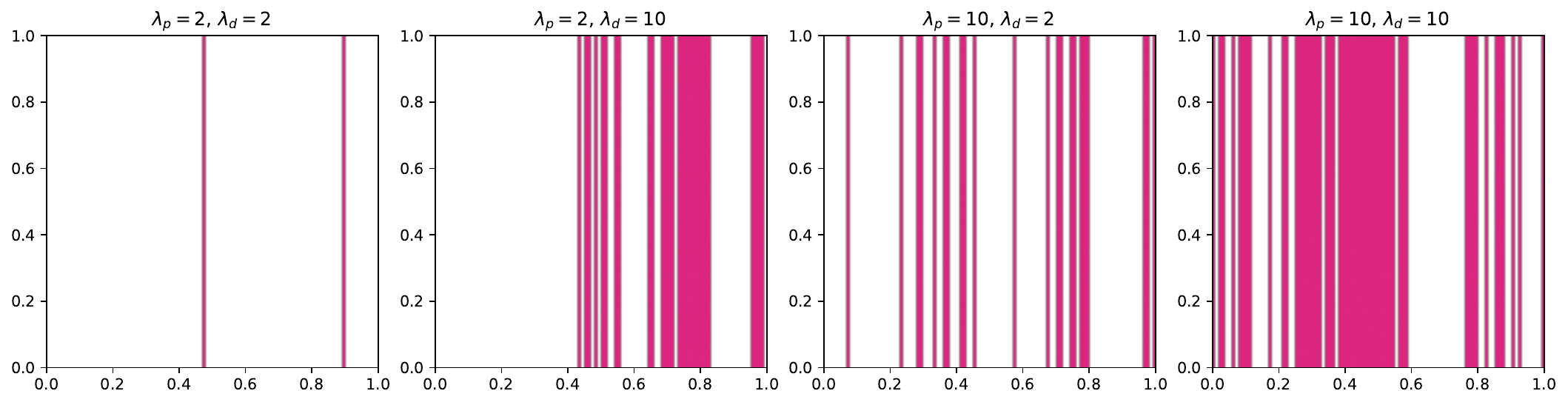}
    \caption{1D spatial $\times$ 1D temporal Matérn Cluster Mask over different parameters}
    \label{fig:poisson_mask_1D}

    \centering
    \includegraphics[width=1\linewidth]{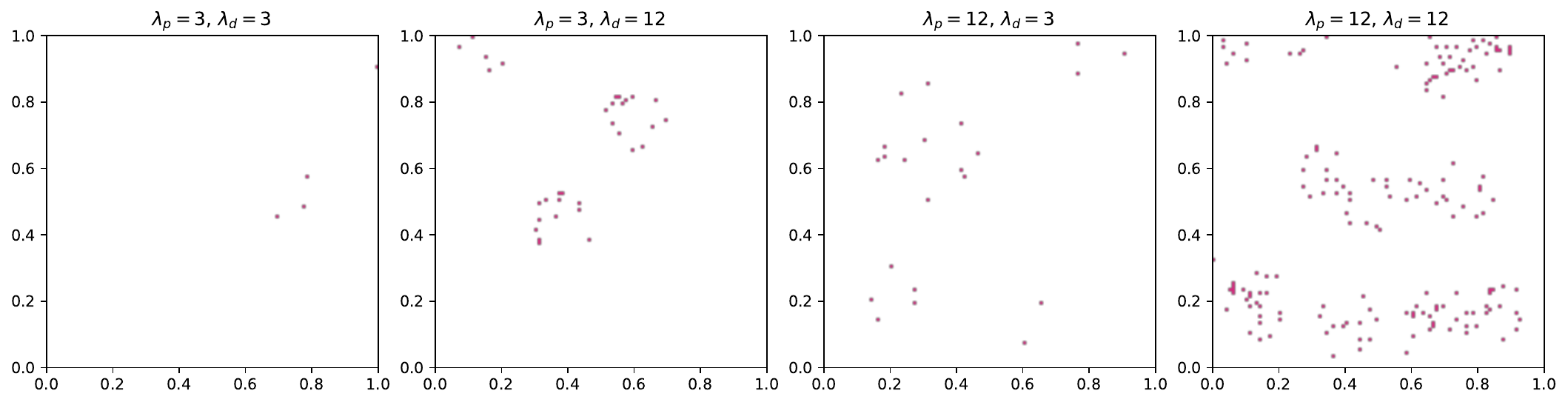}
    \caption{2D spatial Matérn Cluster Mask over different parameters}
    \label{fig:poisson_mask_2D}
\end{figure}

The spatial locations are drawn from a Matérn cluster process following \cite{SainsburyDale2025}. Specifically, the process first distributes a random number of parent points across the domain $\mathcal{D}$, and subsequently generates a random number of daughter points around each parent. The point pattern is then converted into a gridded binary mask following the procedure described in paragraph \ref{subsection:simulation}.
The complete procedure is described in Algorithm \ref{alg:matern}.
As illustrated in Figure \ref{fig:poisson_mask_1D} and Figure \ref{fig:poisson_mask_2D}, this method accommodates a wide range of sampling patterns, varying from entirely sparse observations to more clustered observations. A large number of parent points $\tilde{k}_p$ with a low number of daughter points results in a more dispersed spatial distribution of observations. Conversely, increasing the number of daughter points $\tilde{k}_d$ causes the data to form dense clusters, given that the radius $R$ around the parent points remains fixed.

\vspace{1cm}

\begin{algorithm}[H]
\caption{Matérn Cluster Mask Generation}
\label{alg:matern}

\KwIn{Parent intensity $\lambda_p$, daughter intensity $\lambda_d$, cluster radius $R$}
\KwOut{Binary mask $\boldsymbol{\eta} \in \{0,1\}^K$}

Sample $\tilde{k}_p \sim \mathrm{Poisson}(\lambda_p)$\;

Place parent points
$p_1,\dots,p_{\tilde{k}_p}$ according to a
uniform distribution on $\cal D$\;

\For{$i = 1$ \KwTo $\tilde{k}_p$}{

    Sample $\tilde{k}_d \sim \mathrm{Poisson}(\lambda_d)$\;

    Place daughter points
    $d_1,\dots,d_{\tilde{k}_d}$
    according to a uniform distribution inside the ball of radius $R$
    centered at $p_i$\;
}

Define the binary mask
$\boldsymbol{\eta} \in \{0,1\}^K$
such that grid cells containing at least one point are set to $1$\;

\end{algorithm}

\subsubsection{Neural Networks}

Regardless the batch size, each input to the neural networks has a dimension of $(2, G \times T)$, where the dimension $2$ arises from the concatenation of the data with the mask $\boldsymbol{\eta}$.
We denote the hidden dimension used throughout the entire neural network as $H$ and we set it to $H = 72$. The number of parameters to estimate is equal to $\dim(\Theta) = 5$. 

NBE and NRE neural architectures are constructed in very similar ways as already shown in Figures \ref{fig:architecture_nbe} and \ref{fig:architecture_nre}. The networks can be decomposed into two distinct subnetworks: the first summarizes the data into a summary vector, while the second maps this vector to either the parameter space or a categorical class, depending on the specific task. The embedding network, described in Table \ref{tab:network_arch_common}, first includes convolutional layers (Conv2d) that restrict the number of inputs needed while preserving the strong correlation between the different dimensions \citep{Goodfellow-et-al-2016}. The second component of this network are the max pooling layers. They reduce the dimensionality of the data for the subsequent layers while only extracting the most important features. Given that our data may contain many zeros, this aspect becomes particularly important. The final layer of the CNN, called AdaptiveMaxPool, transforms all the feature maps produced by the convolutions into a single summary vector.

The mapping network for the NBE framework, detailed in Table~\ref{tab:network_arch_nbe}, consists of a standard Multi-Layer Perceptron (MLP). To estimate the $0.05$ and $0.95$ quantiles, we employ two smaller replicates of the complete network architecture (CNN + MLP) configured with $H = 36$. In contrast, for the NRE framework shown in Table~\ref{tab:network_arch_nre}, the parameters $\boldsymbol{\theta}$ are concatenated with the summary statistics immediately following the AdaptiveMaxPool layer. A MLP brings this extended summary statistic to a single output which undergoes a sigmoid layer before being sent to the BCE loss function.

\begin{longtable}[]{@{}lccccr@{}}
\caption{CNN embedding architecture, common to NBE and NRE.}\label{tab:network_arch_common}\tabularnewline
\toprule\noalign{}
Layer Type & Output Shape & Kernel & Padding & Activation & Parameters \\
\midrule\noalign{}
\endfirsthead
\toprule\noalign{}
Layer Type & Output Shape & Kernel & Padding & Activation & Parameters \\
\midrule\noalign{}
\endhead
\bottomrule\noalign{}
\endlastfoot
Conv2d (1) & $(H/2, 97 \times 96)$ & $5 \times 8$ & $2 \times 3$ & ReLU & $2{,}916$ \\
Conv2d (2) & $(H/2, 97 \times 96)$ & $5 \times 5$ & $2 \times 2$ & ReLU & $32{,}436$ \\
MaxPooling & $(H/2, 48 \times 48)$ & $2 \times 2$ & -- & -- & $0$ \\
Conv2d (3) & $(H, 48 \times 48)$ & $3 \times 5$ & $1 \times 2$ & ReLU & $38{,}952$ \\
Conv2d (4) & $(H, 48 \times 48)$ & $3 \times 3$ & $1 \times 1$ & ReLU & $46{,}728$ \\
AdaptiveMaxPool & $H$ & -- & -- & -- & $0$ \\
\end{longtable}

\begin{longtable}[]{@{}lccr@{}}
\caption{NBE mapping neural network architecture.}\label{tab:network_arch_nbe}\tabularnewline
\toprule\noalign{}
Layer Type & Output Shape & Activation & Parameters \\
\midrule\noalign{}
\endfirsthead
\toprule\noalign{}
Layer Type & Output Shape & Activation & Parameters \\
\midrule\noalign{}
\endhead
\bottomrule\noalign{}
\endlastfoot
Linear (1) & $H$ & ReLU & $5{,}256$ \\
Linear (2) & $\dim({\Theta})$ & -- & $219$ \\
\end{longtable}

\begin{longtable}[]{@{}lccr@{}}
\caption{NRE mapping neural network architecture.}\label{tab:network_arch_nre}\tabularnewline
\toprule\noalign{}
Layer Type & Output Shape & Activation & Parameters \\
\midrule\noalign{}
\endfirsthead
\toprule\noalign{}
Layer Type & Output Shape & Activation & Parameters \\
\midrule\noalign{}
\endhead
\bottomrule\noalign{}
\endlastfoot
\textit{Concatenation} & $H + \dim{\Theta}$ & -- & $0$ \\
Linear (1) & $H$ & ReLU & $5{,}472$ \\
Linear (2) & $1$ & -- & $73$ \\
\end{longtable}

\subsection{Results}


To demonstrate the efficiency of our method, we evaluate its performance within a fully simulated framework. We conduct our studies using a workstation with an Intel Xeon W-2223 3.60GHz CPU (4 cores, 8 threads), 64 GB of CPU RAM, and a single NVIDIA RTX A5000 GPU with 24 GB of GPU RAM. All the results of the paper are reproducible from the code available on \href{https://github.com/DarenXL/Neural_ST_inference}{GitHub}.

We first sample 10000 different sets of parameters $(v, \rho_t, \rho_s,\sigma)$ for the training dataset and 2000 others for the testing dataset. For each set of parameters, we simulate one corresponding spatio-temporal Gaussian random field (\ref{eq:model_matern}) on a complete grid of size $100 \times 100$. A random independent gaussian noise $\mathcal{N}(0, \tau^2)$ is added to mimic an observation noise with $\tau$ randomly drawn from its prior uniform distribution.
During both training and testing phases, the spatial locations are sampled according to Algorithm~\ref{alg:matern}. Values at locations hidden by the mask are set to zero. Consequently, a single complete data grid simulation can yield several distinct scattered simulations by simply varying the mask. In the training dataset, the missing data ratio is highly variable to ensure that the network adapts to a wide range of observational scenarios. In contrast, we choose a high missing data ratio of 86\% during the testing phase to evaluate the models under a challenging scenario. 

\begin{figure}
    \centering
    \includegraphics[width=0.95\linewidth, trim={0 0 0 .8cm}, clip]{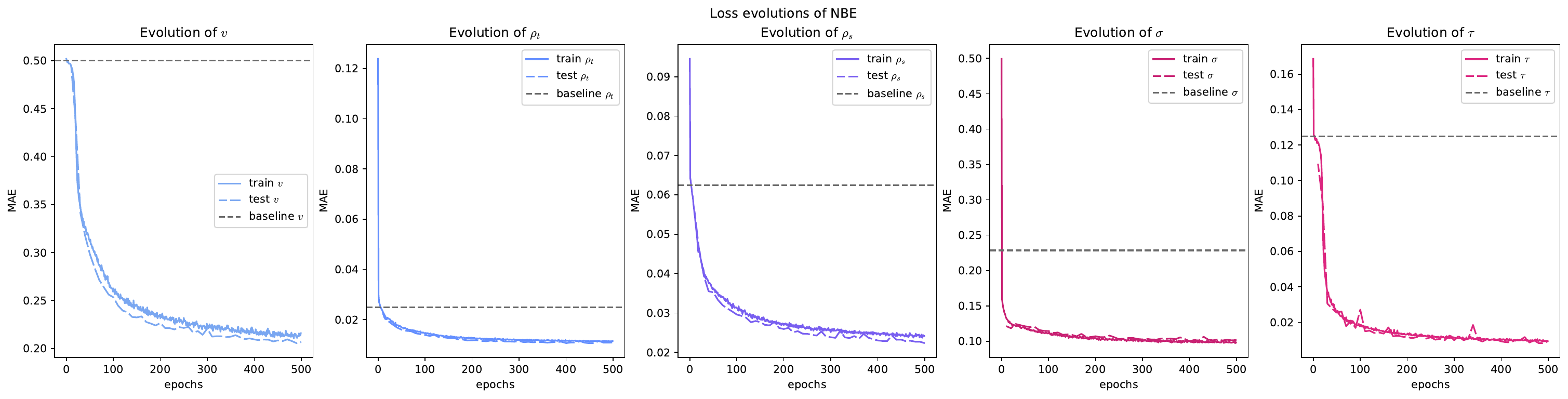}
    \caption{Evolution of the NBE's MAE loss on the training (solid lines) and test (dashed lines) datasets over 500 epochs. The gray dashed lines represent the baseline for each parameter, i.e. the MAE between the real parameters and the mean of their prior uniform distribution.}
    \label{fig:train_nbe}
\end{figure}

\begin{figure}
    \centering
    \includegraphics[width=.7\linewidth]{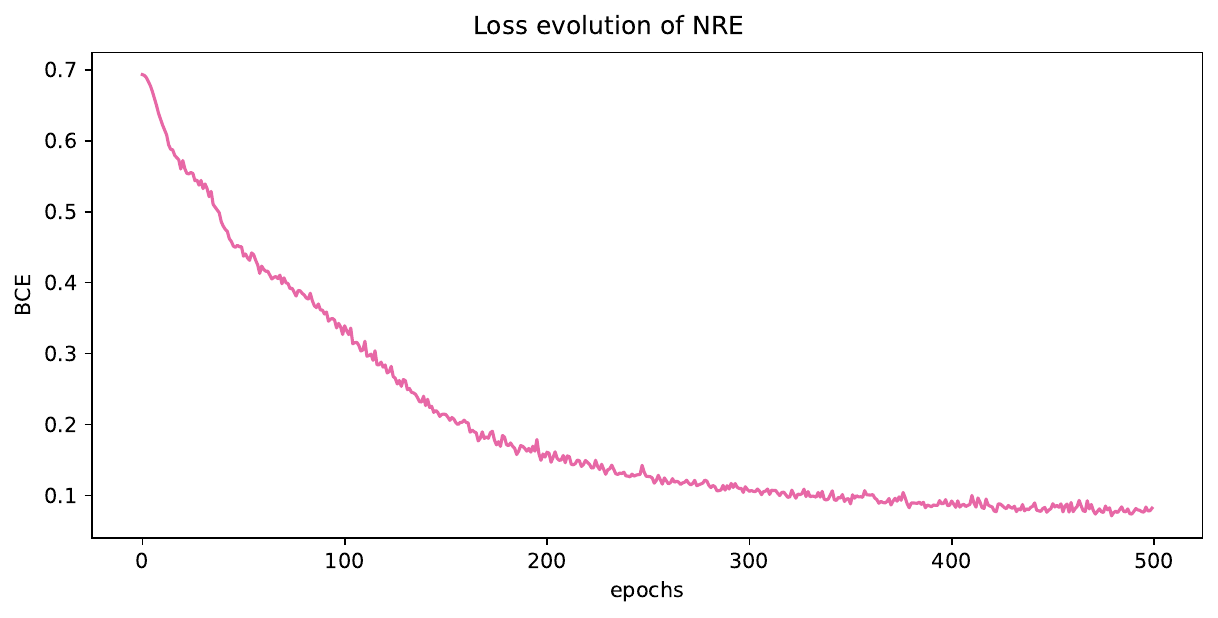}
    \caption{Evolution of the NBE's BCE loss on the training dataset over 500 epochs.}
    \label{fig:train_nre}
\end{figure}

The neural networks are described in the previous section. 
They are trained separately for 500 epochs using the Adam optimizer \citep{kingma2014adam}, with learning rates set to $0.001$ for NBE and $0.00005$ for NRE. The evolutions of the losses are shown in Figure \ref{fig:train_nbe} and Figure \ref{fig:train_nre}. In the NBE case, the baseline shown is the MAE between the real parameters and the mean of the prior distributions. Moreover, NBE's network is trained along smaller side networks to target the quantiles $0.05$ and $0.95$.
The curves appear to show a convergence in both train and test sets.

\begin{figure}
    \centering
    \includegraphics[width=1\linewidth, trim={0 0 0 1cm}, clip]{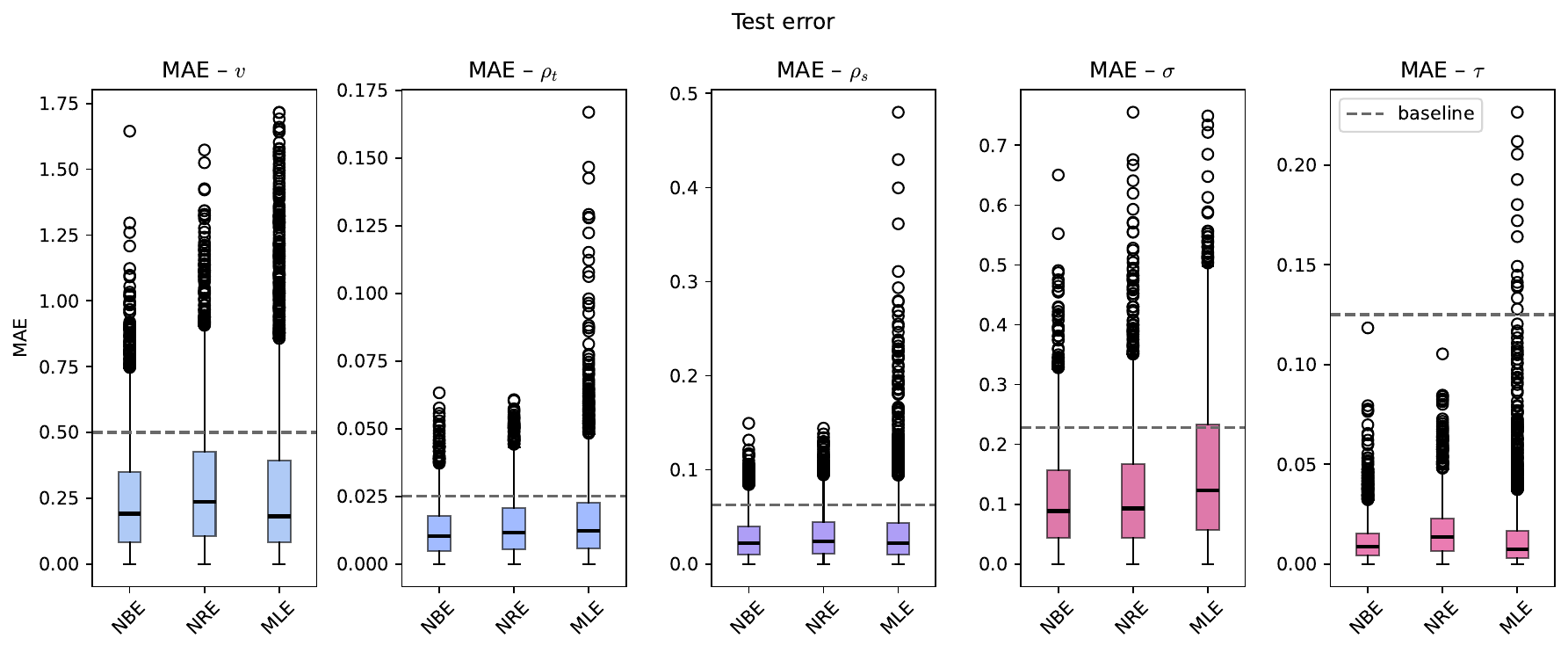}
    \caption{MAE scores for the NBE and NRE methods and MLE for each parameter with a missing ratio of ~86\% in the data. The baseline (gray dashed lines) for each parameter is the MAE between the real parameters and the mean of the prior uniform distributions. NRE's point estimate is the median of the posterior distribution obtained via MCMC sampling. MLE is computed with the Adam optimizer (learning rate of 0.05 and 600 iterations) with 300 points chosen randomly.}
    \label{fig:box_scores}
\end{figure}

Figure~\ref{fig:box_scores} shows the MAE scores computed on the test set for both NBE and NRE methods in comparison to the MLE as baseline. To obtain the MLE in a reasonable computational time, for each element of the test dataset, we choose randomly 300 observation points. The maximization is done with the Adam optimizer with 0.05 as learning rate over 600 iterations, using \texttt{GPyTorch} \citep{gardner2018gpytorch} accelerated with the package \texttt{KeOps} \citep{JMLR:v22:20-275}, as proposed in \citet{romary2026deep}. For the NRE framework, the point estimate is defined as the median of the posterior distribution obtained via MCMC sampling. According to the box plots, the NBE appears to perform slightly better than NRE. Nevertheless, both methods yield competitive results, as highlighted by the fact that all boxes (and thus their third quartiles) fall below the baselines. These baselines correspond to the MAE between the real parameters and the mean of their uniform prior distribution, already shown in Figure~\ref{fig:train_nbe}. 
As for Uncertainty quantification (UQ), Table~\ref{tab:quantiles} displays the coverage probability, representing the ratio of true parameters that fall within the estimated $0.05$ and $0.95$ quantiles, where the target is $0.90$ for each parameter. For both methods, quantiles are close to the target coverage. On the contrary, the MLE method doesn't provide any confidence interval. 

\begin{longtable}[]{@{}l  lllll@{}}
\caption{Coverage probability for each estimated parameter. The target is $0.90$ for each parameter.}\label{tab:quantiles}\tabularnewline
\toprule\noalign{}
Method & $v$ & $\rho_t$ & $\rho_s$ & $\sigma$ & $\tau$ \\
\midrule\noalign{}
\endfirsthead
\toprule\noalign{}
Method & $v$ & $\rho_t$ & $\rho_s$ & $\sigma$ & $\tau$ \\
\midrule\noalign{}
\endhead
\bottomrule\noalign{}
\endlastfoot
NBE & 0.918 & 0.897 & 0.902 & 0.879 & 0.878 \\
NRE & 0.900 & 0.915 & 0.892 & 0.898 & 0.920 \\
\end{longtable}

\begin{figure}[t]
    \centering
    \includegraphics[width=.7\linewidth, trim={0 0 0 0.75cm}, clip]{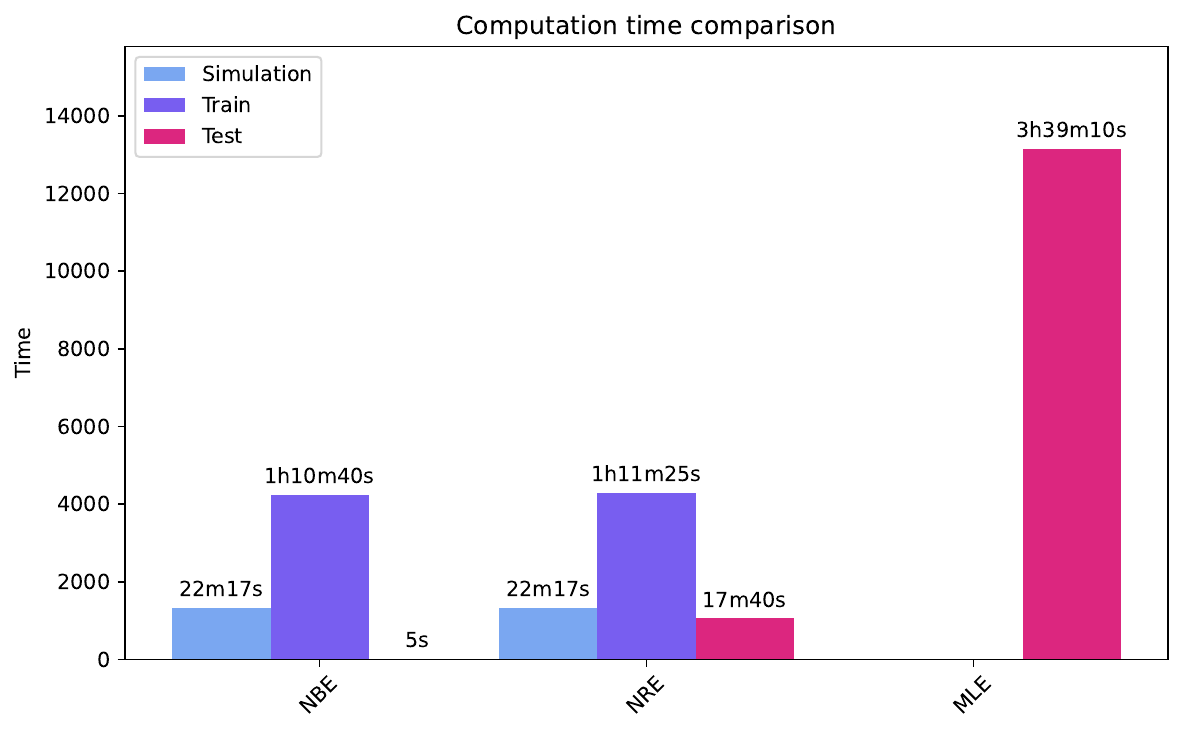}
    \caption{Simulation, training and test times for the NBE and NRE methods and MLE. Simulation time is the time taken by \texttt{gstlearn} to simulate the training dataset of 5000 complete grids (on CPU). The NBE training time includes networks dedicated to quantile estimation. Test time for NRE is an MCMC run of the NUTS algorithm on 3 chains with 500 warm-up steps and 2000 samples. MLE is computed with the Adam optimizer (learning rate of 0.05 and 600 iterations) with 300 points chosen randomly.}
    \label{fig:times_results}
\end{figure}

Another point of comparison concerns the computation times taken by each method.
As illustrated in Figure~\ref{fig:times_results}, NBE outperforms NRE also in terms of overall computational efficiency. The simulation time is identical for both approaches since they share the same training dataset generation process. The training time for NBE is slightly longer, driven by the need to train two additional networks dedicated to quantile estimation. The training time for NRE is substantial due to the doubling of its effective dataset size, where the second half is constructed by shuffling the couples of the first half.
The most significant difference arises in test time: NBE provides instantaneous direct inference, whereas NRE requires MCMC sampling to approximate the full posterior distribution and its corresponding quantiles. Even though the NUTS algorithm minimizes the need for manual hyperparameter tuning, choices like the number of chains and their lengths impact the computational time and the accuracy of the posterior approximation. With the current setting, the total test time for $2000$ simulations is lower than the training time, and the results seem reasonably satisfactory. Despite the high processing speed of the computation of the MLE, the lack of amortization of this method is evident on a large test set.

\begin{figure}
    \centering
    \includegraphics[width=1\linewidth, trim={0 0 0 1cm}, clip]{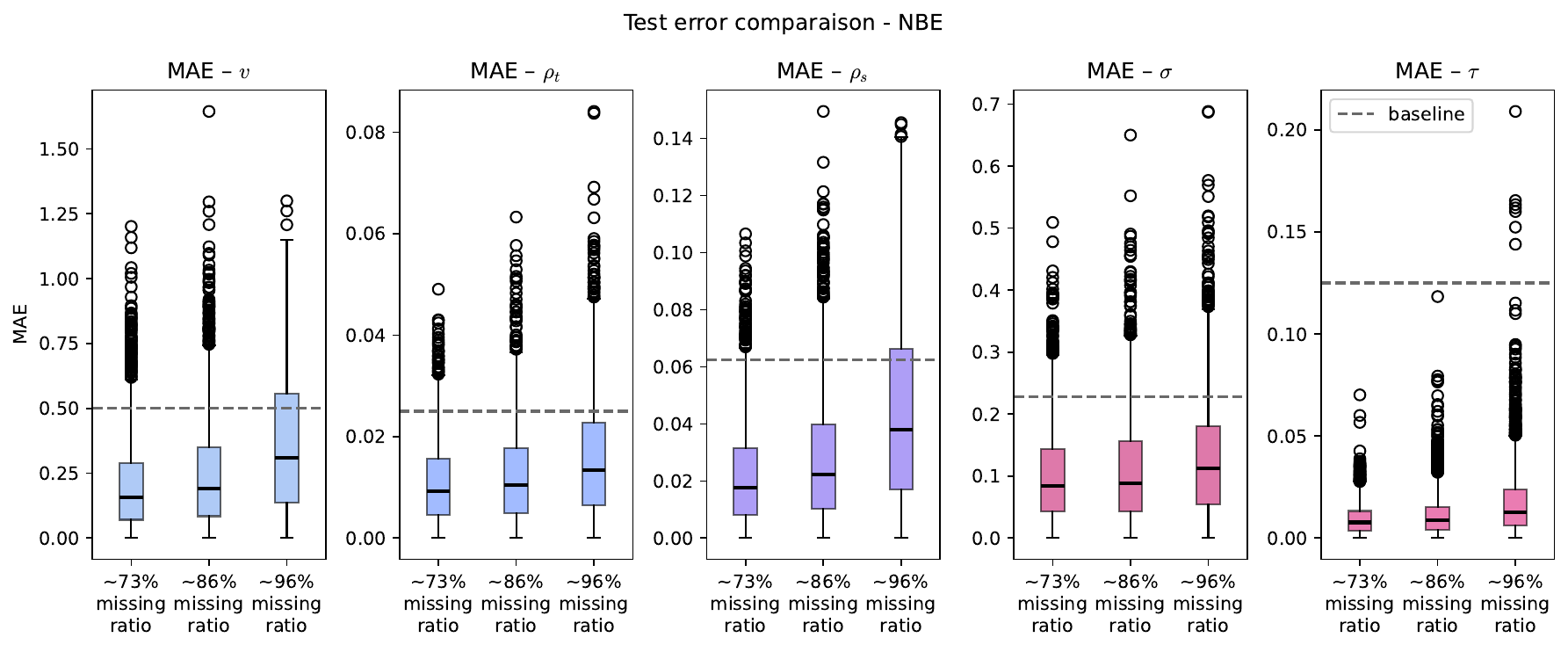}
    \caption{MAE scores for the NBE method for each parameter on different missing ratios. The baseline (gray dashed lines) for each parameter is the MAE between the real parameters and the mean of the prior uniform distributions.}
    \label{fig:box_scores_NBE}
\end{figure}

\begin{figure}
    \centering
    \includegraphics[width=1\linewidth, trim={0 0 0 1cm}, clip]{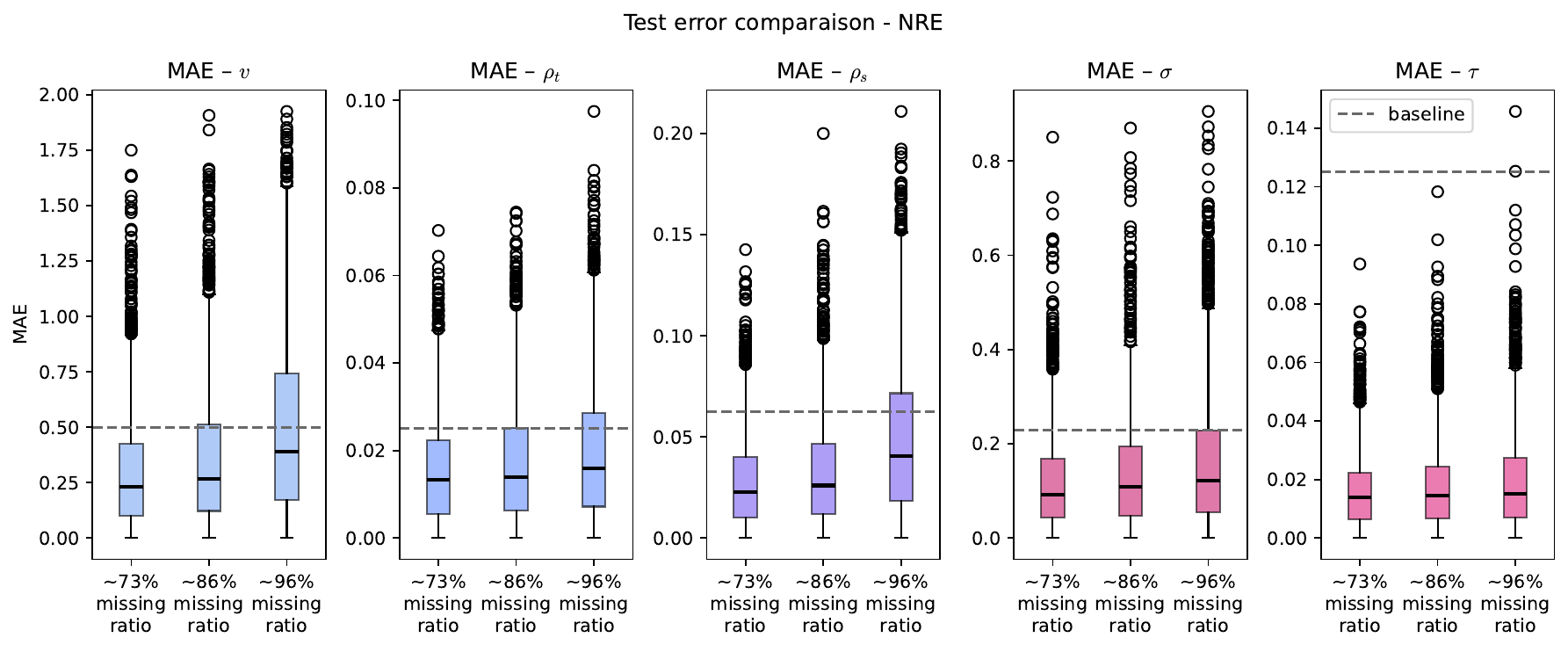}
    \caption{MAE scores for the NRE method for each parameter on different missing ratios. The baseline (gray dashed lines) for each parameter is the MAE between the real parameters and the mean of the prior uniform distributions. The point estimate is obtained via gradient descent optimization.}
    \label{fig:box_scores_NRE}
\end{figure}

A notable trend regarding the performance metrics emerges when examining Figures~\ref{fig:box_scores_NBE} and \ref{fig:box_scores_NRE}. The former displays the NBE scores across three distinct missing data ratios: 73\%, 86\%, and 96\%. As anticipated, a higher ratio of missing data degrades performance, resulting in higher MAE scores. The 96\% missing ratio scenario, also highlighted in Figure~\ref{fig:box_scores}, is particularly challenging, because a significant portion of the test cases contain only a single temporal observation, making the estimation highly uncertain and leading to worse scores. Figure~\ref{fig:box_scores_NRE} illustrates the corresponding results for the NRE method, where point estimations are derived via gradient descent optimization rather than MCMC sampling. Even though this optimization approach proves less effective than MCMC under extreme data scarcity (the 96\% missing ratio), the scores remain below the baseline. Additionally, gradient descent optimization provides a major speed advantage in the inference part, with a computation time of 4m42s compared to the 17m40s from the MCMC.

\begin{figure}
    \centering
    \includegraphics[width=1\linewidth]{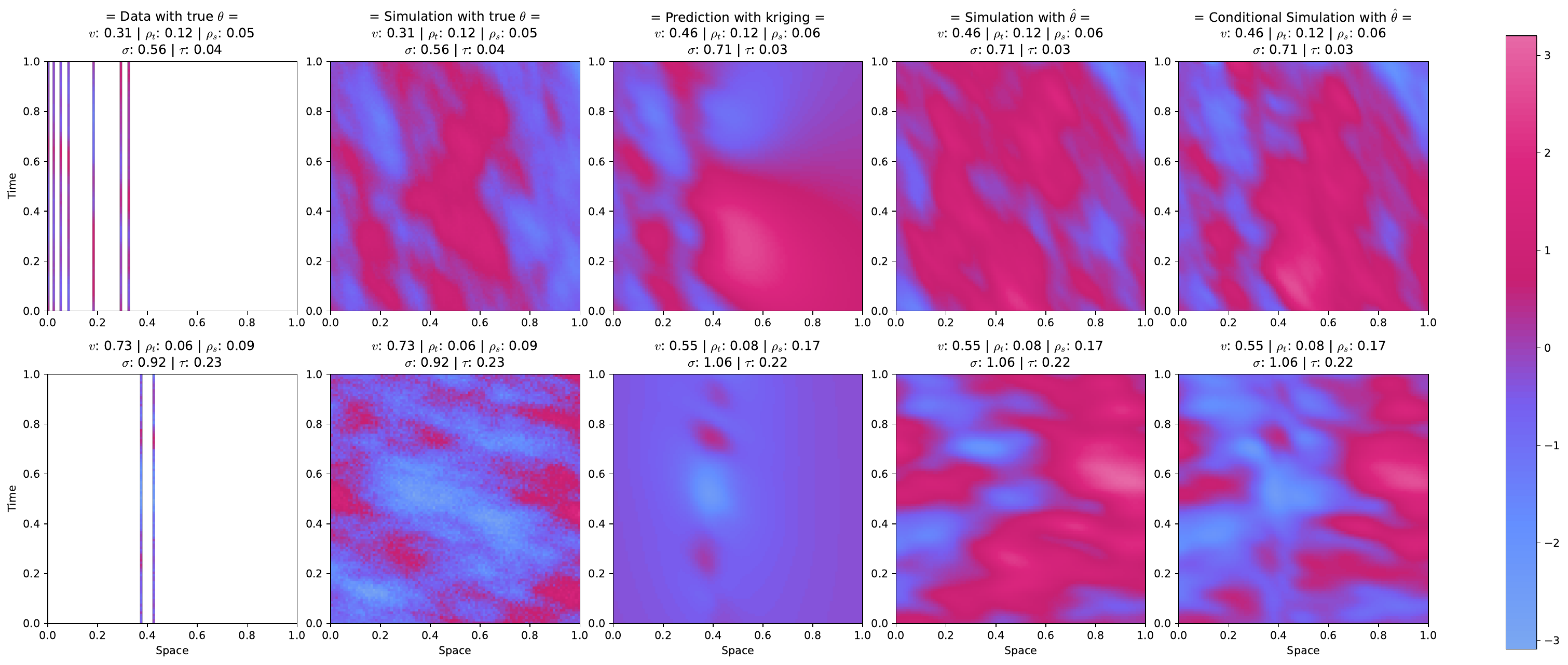}
    \caption{Illustration of prediction results in two scenarios. Columns represents, from left to right: the observations, the true unobserved Gaussian field, the kriging interpolation, one unconditional simulation and one conditional simulation. True parameters are displayed above the first two columns, while estimated parameters are reported above the remaining three columns.}
    \label{fig:visualisation}
\end{figure}

With the estimated parameters by NBE, we perform prediction over the complete spatio-temporal grid.
Figure~\ref{fig:visualisation} illustrates two examples of prediction. The first column corresponds to the observations and the second one to the true unobserved Gaussian field. The third, fourth, and fifth columns display the kriging interpolation, one unconditional simulation, and one conditional simulation, respectively, all generated using the estimated parameters. The inherent smoothing effect of kriging is clearly illustrated here, becoming increasingly pronounced in regions far from any available observations. While the unconditional simulation generates a field with a realistic spatio-temporal structure, it does not incorporate the observations from the first column. In contrast, the conditional simulation successfully reproduces realistic spatio-temporal structures while strictly honoring the observed values.

To provide a benchmark across methods, kriging predictions are evaluated at all missing locations for each simulation in the test set. Table~\ref{tab:prediction_metric} compares the results of NBE, NRE, and MLE across three distinct performance metrics : mean squared error (MSE), mean absolute error (MAE) and Negative Log Predictive Density (NLPD). Notably, the NBE approach consistently yields the best results, marginally outperforming all competing methods across every metric. Negative values for the NLPD metric suggest that our models are not only accurate, but also confident in their predictions.

\begin{longtable}[]{@{}lccc@{}}
\caption{Comparison of prediction metrics across models. For all reported metrics, lower values represent better performance. The Negative Log Predictive Density (NLPD) exploits the theoretical distributions of the spatio-temporal models to penalize overconfident and underconfident predictions.} 
\label{tab:prediction_metric}\tabularnewline
\toprule\noalign{}
Method & MSE & MAE & NLPD \\
\midrule\noalign{}
\endfirsthead
\toprule\noalign{}
Method & MSE & MAE & NLPD \\
\midrule\noalign{}
\endhead
\bottomrule\noalign{}
\endlastfoot
NBE & $0.408$ & $0.447$ & $-0.097$ \\
NRE & $0.416$ & $0.451$ & $-0.091$ \\
MLE & $0.419$ & $0.453$ & $-0.090$ \\
\end{longtable}

These results are promising, demonstrating that our approach enables fast and reliable parameter estimation for complex spatio-temporal models. Furthermore, the CNN architecture effectively handles missing observations without compromising accuracy.

\section{Application}  \label{section:applications}

To demonstrate the effectiveness of our approach, we train a neural network tailored to three-dimensional Gaussian Random Fields (two spatial dimensions and one temporal dimension) for a real-world application. Specifically, we leverage the amortized nature of simulation-based inference (SBI) to efficiently estimate parameters from spatio-temporal solar radiation data.

\subsection{Solar radiation data}

The selected dataset comes from the HOPE campaign \citep{macke2017hd} which records the Global Horizontal Irradiance (GHI) in West Germany during three months from April 2 to July 2, 2013. The GHI (W/m²) represents the amount of light —both direct beam and diffuse sky radiation— hitting a horizontal surface, parallel to the ground. 
The GHI is captured every minute from 99 measuring stations spread over a (16 × 16) km² region. 


The preprocessing phase involved extracting the deterministic component from the data. Following \cite{oumbe2014decoupling}, the GHI ($G$) is factorized into the product of a deterministic clear-sky model ($G_c$), representing direct solar radiation, and a stochastic term (the clear-sky index $K$), which accounts for cloud attenuation: 
$$G = G_c K.$$
Since the clear-sky model, capturing diurnal solar cycles, can be reliably computed \citep{gschwind2019improving}, the subsequent analysis focuses exclusively on modeling the stochastic component $K$.

\begin{figure}[htbp]
    \centering
    
    \begin{subfigure}[b]{0.48\textwidth}
        \centering
        \includegraphics[width=\linewidth]{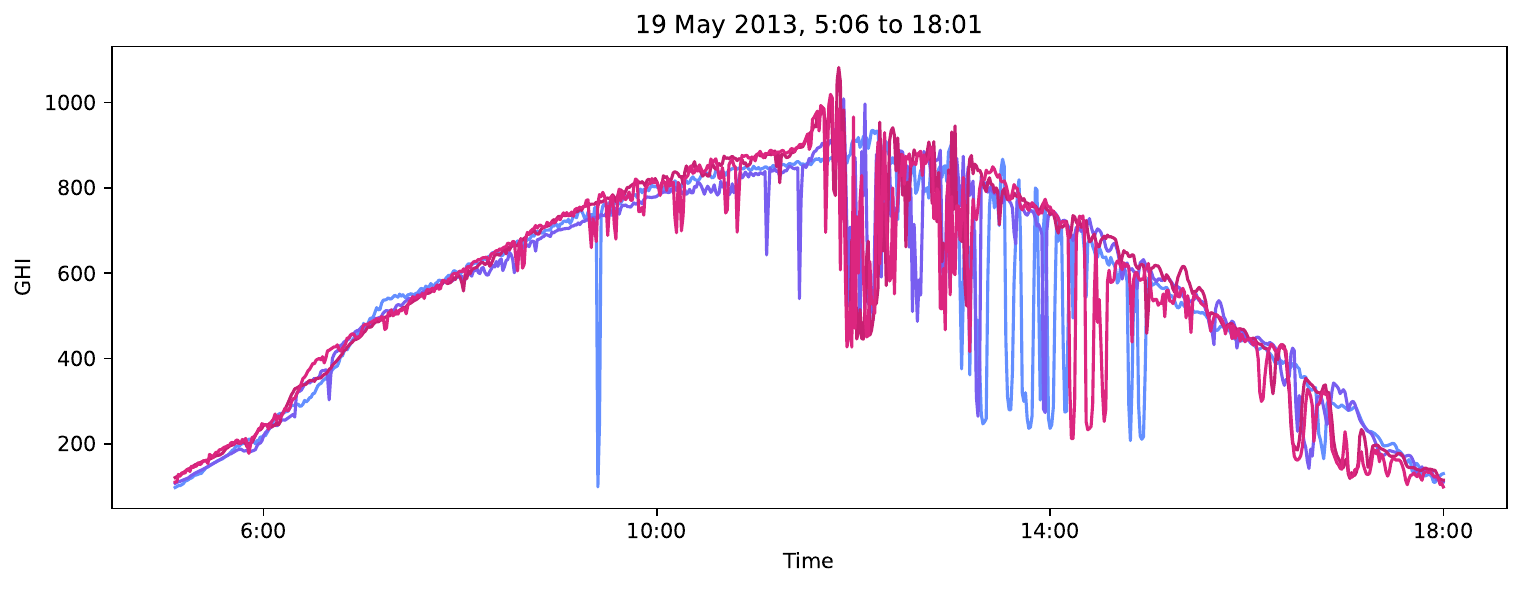}
        \caption{GHI $G$ on May 19, 2013.}
        \label{fig:19may}
    \end{subfigure}
    \hfill
    \begin{subfigure}[b]{0.48\textwidth}
        \centering
        \includegraphics[width=\linewidth]{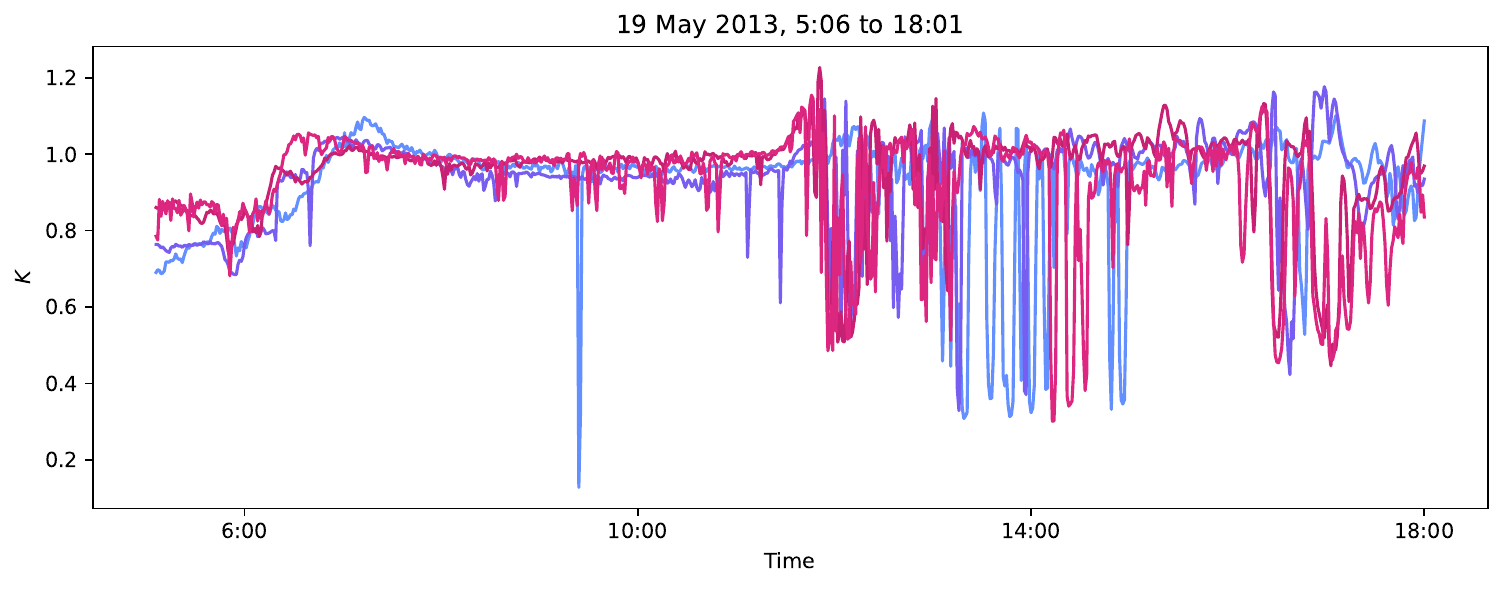}
        \caption{Clear-sky index $K$ on May 19, 2013.}
        \label{fig:19may_nodeter}
    \end{subfigure}

    \vspace{0.8em} 

    \begin{subfigure}[b]{0.48\textwidth}
        \centering
        \includegraphics[width=\linewidth]{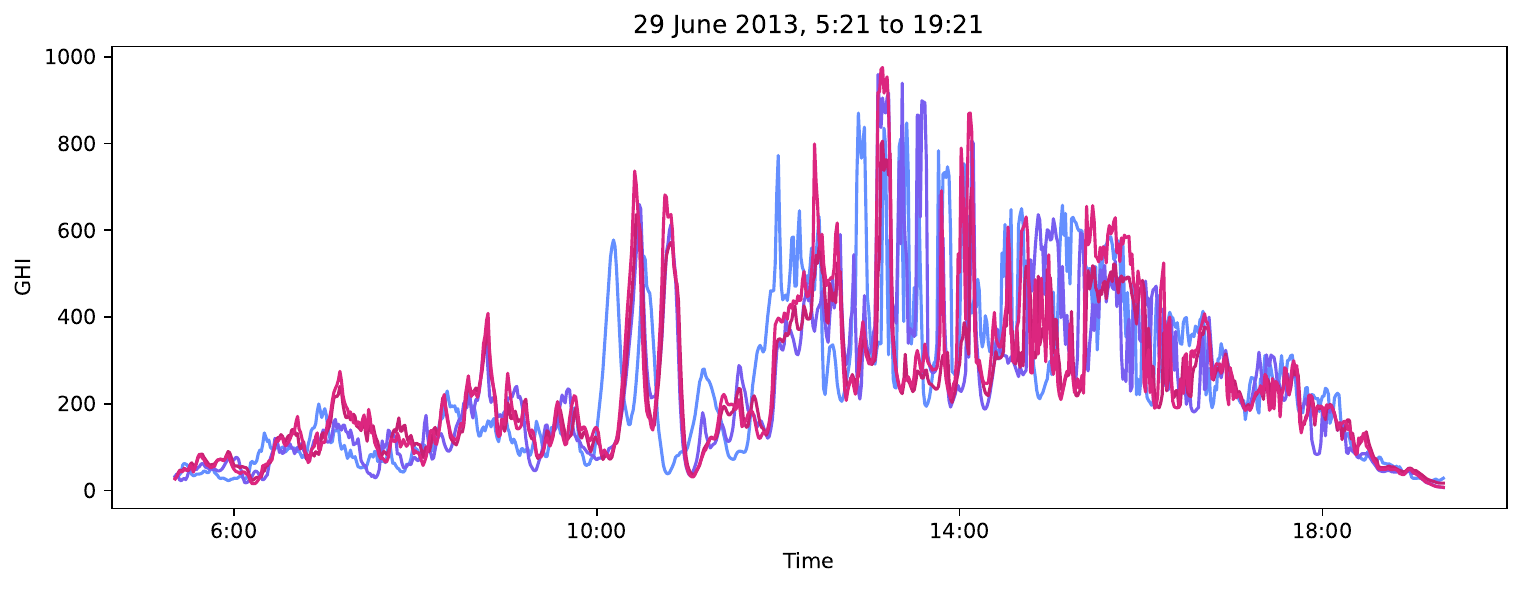}
        \caption{GHI $G$ on June 29, 2013.}
        \label{fig:29june}
    \end{subfigure}
    \hfill
    \begin{subfigure}[b]{0.48\textwidth}
        \centering
        \includegraphics[width=\linewidth]{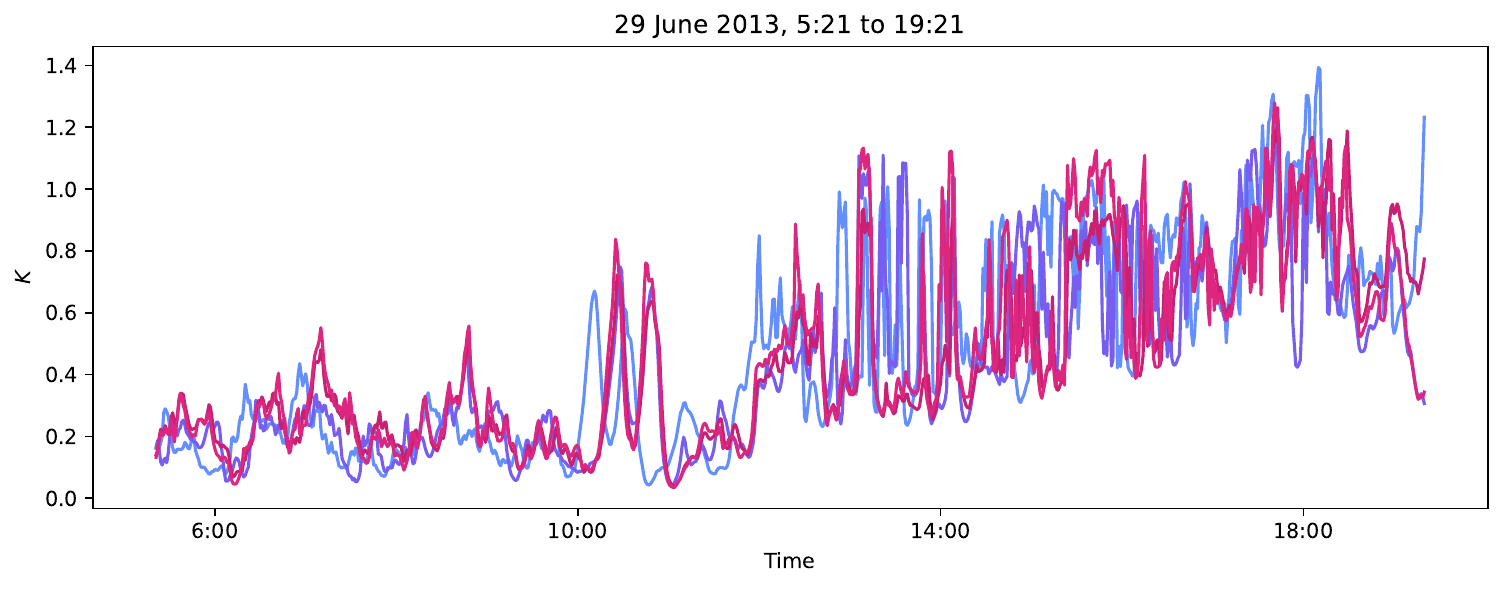}
        \caption{Clear-sky index $K$ on June 29, 2013.}
        \label{fig:29june_nodeter}
    \end{subfigure}

    \caption{Illustration of GHI $G$ (left) and clear-sky index $K$ (right) on a sunny (May 19, 2013) and a cloudy day (June 29, 2013).}
    \label{fig:comparaison_dates}
\end{figure}

Figure~\ref{fig:comparaison_dates} illustrates both GHI $G$ and clear-sky index $K$ on two different days in 2013. The daily evolution of both variables highlights differences in cloud dynamics, contrasting a sunny day (May 19, 2013), characterized by very few clouds, with a cloudy day (June 29, 2013), marked by an attenuation of GHI throughout the day.

Thanks to the pronounced presence of cloud cover and wind, June 29 was selected as a case study. Such meteorological conditions are ideal for evaluating our framework, as cloud dynamics play a central role: variations in cloud density drive changes in spatial and temporal ranges, while cloud motion induces advection. After removing sensors with missing values and restricting the time frame to daylight hours, the resulting dataset consists of 840 time steps (from 05:21 to 19:21) across 77 spatial locations.

\subsection{Inference and prediction}

We model the clear-sky index on June, 29 as a spatio-temporal Gaussian random field as in Equation \eqref{eq:model_ST} with a fully non-separable Gneiting-Matérn covariance structure with advection (Equation \eqref{eq:model_matern} with $\gamma=1$). 

\begin{figure}
    \centering
    \includegraphics[width=.7\linewidth]{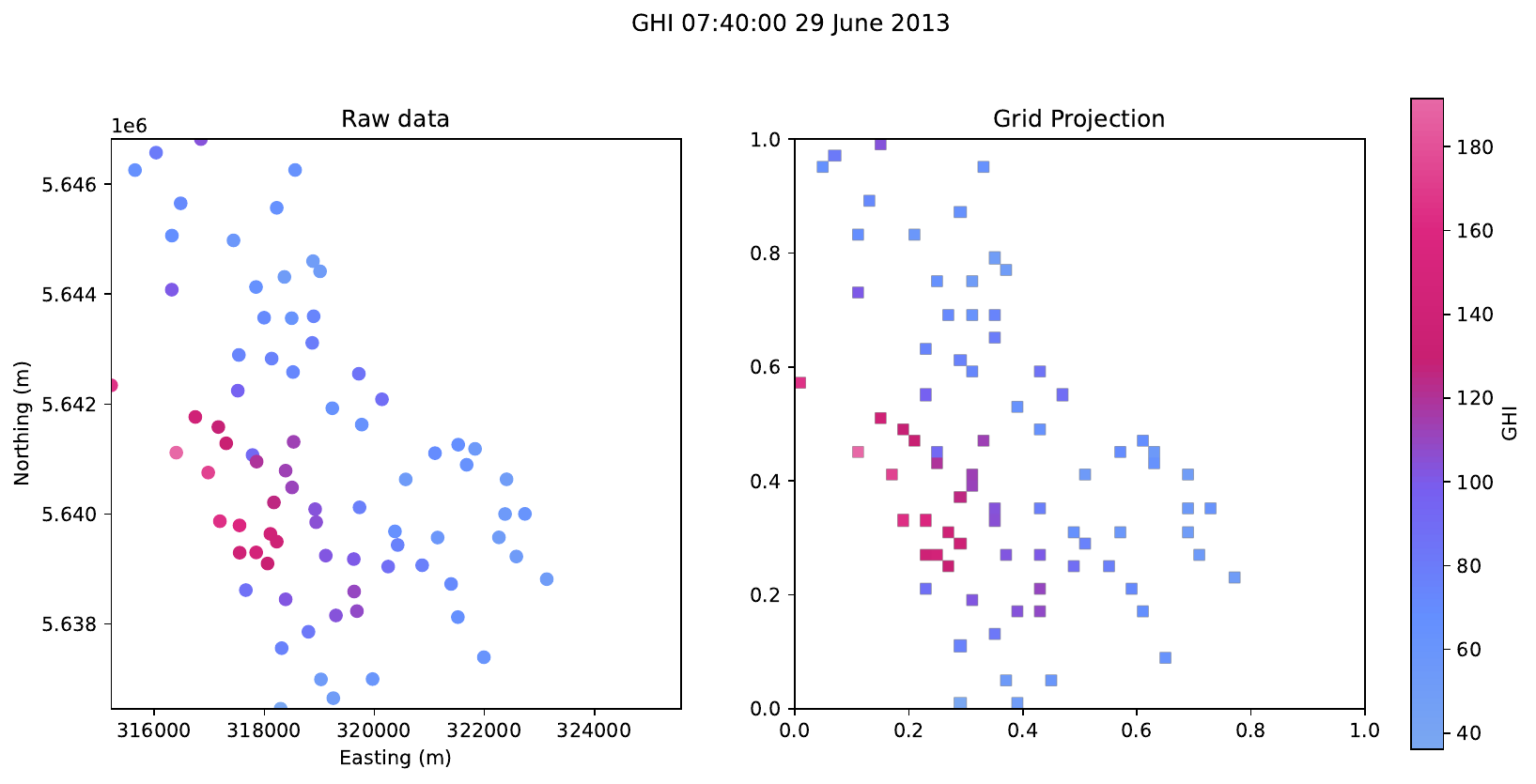}
    \caption{Solar radiation at measured stations (left) and after projection over the spatial grid (right).}
    \label{fig:projection}
\end{figure}

We project the scattered stations onto a $50 \times 50$ spatial grid of size $(320\times 320) \text{m}^2$ per pixel, following the approach introduced in Section \ref{subsection:simulation}.
Figure \ref{fig:projection} illustrates the data distribution following projection.

We adopt a fully amortized framework based on the NBE approach, including uncertainty quantification addressed with dedicated quantile networks. The network was trained on a dataset of $10000$ simulations over a spatio-temporal grid of size $K \times K \times m = 50 \times 50 \times 20$. Consistent with the results in Figure~\ref{fig:box_scores_NBE}, the network successfully infers the spatio-temporal parameters in this three-dimensional setting. The total computational budget required approximately $2$ hours for simulation generation and $17$ hours for training. This network generalizes effectively beyond the solar radiation setting and can be directly applied to parameter inference across diverse spatio-temporal datasets.

We employ the trained NBE model to estimate the vector of spatio-temporal parameters $\boldsymbol{\theta} = (v_1, v_2, \rho_t, \rho_s, \sigma, \tau)$ directly from solar radiation data. 

Accounting for the temporal dimension presents a key challenge when estimating parameters over a 840-minute daytime period. Processing the dataset as a single batch is insufficient, as underlying parameters vary dynamically over time, due to shifts in atmospheric conditions, such as wind speed and direction.

\begin{figure}
    \centering
    \includegraphics[width=1\linewidth]{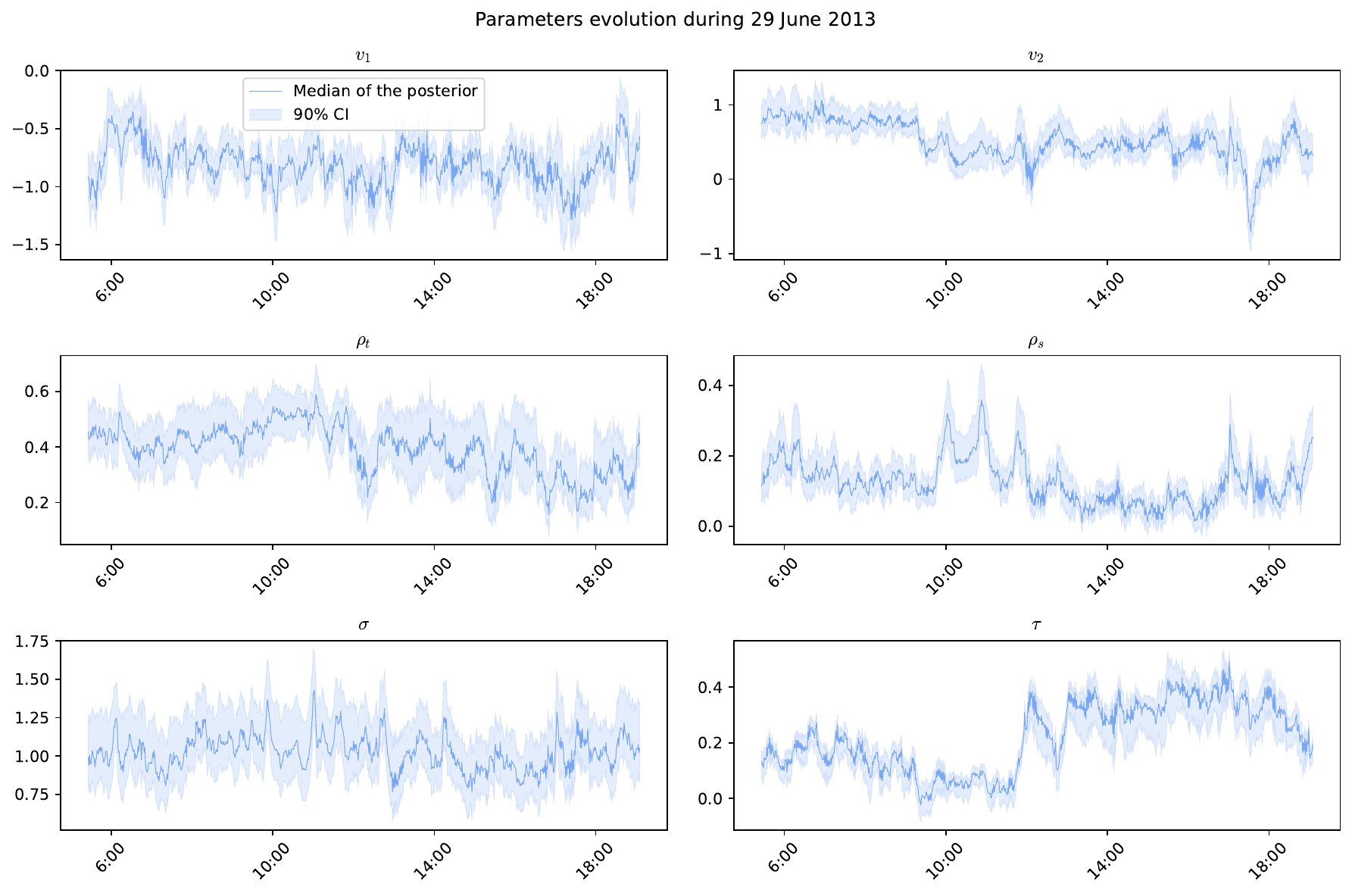}
    \caption{Temporal evolution of inferred parameters $\widehat{\boldsymbol{\theta}}_j$ obtained via 20-minute sliding-window inference on June 29, 2013.}
    \label{fig:window_size}
\end{figure}

To address this temporal non-stationarity, we implement a sliding-window approach using the trained network. Specifically, for each time step $j \in 20, \dots, 840$ (spanning from 05:41 to 19:21), we estimate the parameter vector $\boldsymbol{\theta}_j$ via a forward pass through the NBE network, using the standardized spatio-temporal observations from the preceding 20-minute interval. Figure~\ref{fig:window_size} illustrates the temporal evolution of the estimated median $\widehat{\boldsymbol{\theta}}_j$ along with the $90\%$ confidence interval for June 29, 2013. The results reveal a clear dynamic trend in the covariance parameters throughout the day.

\begin{figure}
    \centering
    \includegraphics[width=0.9\linewidth]{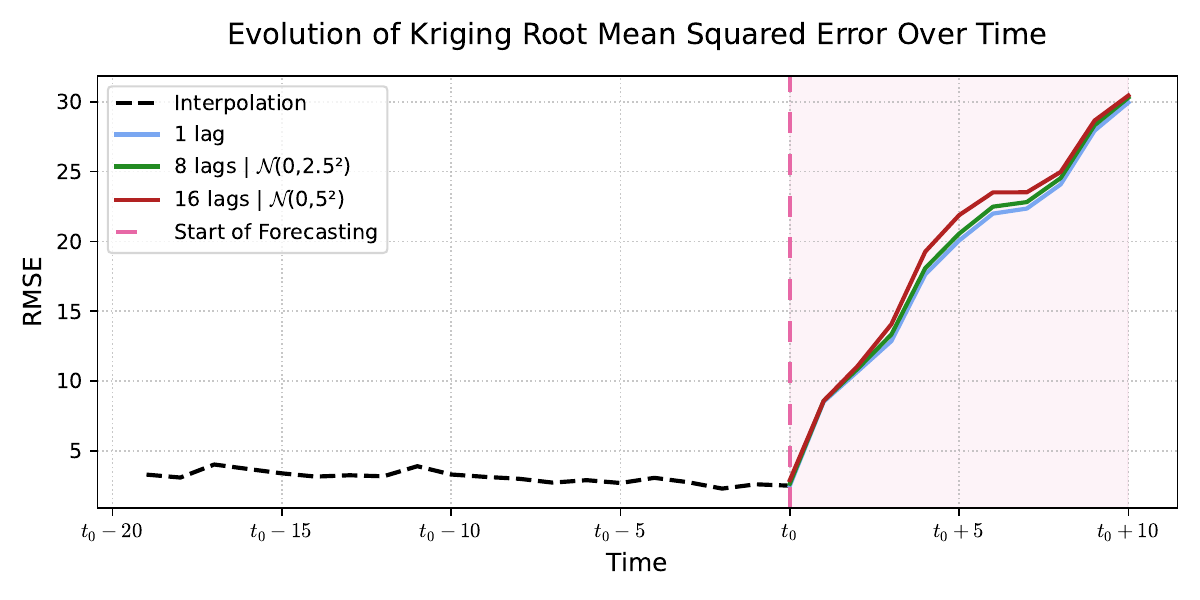}
    \caption{Root Mean Squared Error from interpolation ($[t_0 - 19,t_0]$) and forecasting ($[t_0 + 1,t_0 + 10]$) by kriging on June 29, 2013 (7:21--7:50). The dashed line represents interpolation error, while solid lines denote forecasting error under different parameterizations. Forecasting parameters are smoothed using a half-Gaussian filter with varying time lags and standard deviations, indicated in the legend.} 
    \label{fig:kriging_error_timeseries}
\end{figure}

\begin{figure}[htbp]
    \centering
    
    \begin{subfigure}[b]{1\textwidth}
        \centering
        \includegraphics[width=\linewidth]{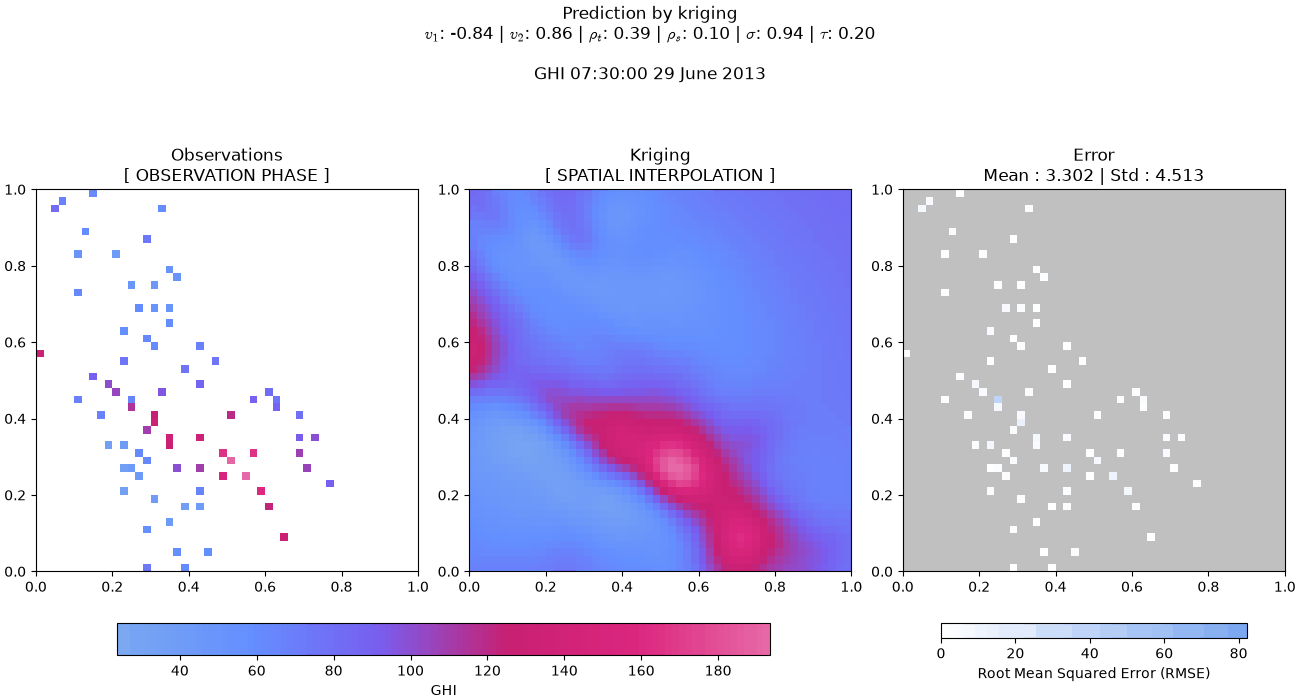}
        \caption{Interpolation at 7:31 on June 29, 2013.}
        \label{fig:3D_krig_interpolation}
    \end{subfigure}

    \vspace{0.8em} 
    
    \begin{subfigure}[b]{1\textwidth}
        \centering
        \includegraphics[width=\linewidth]{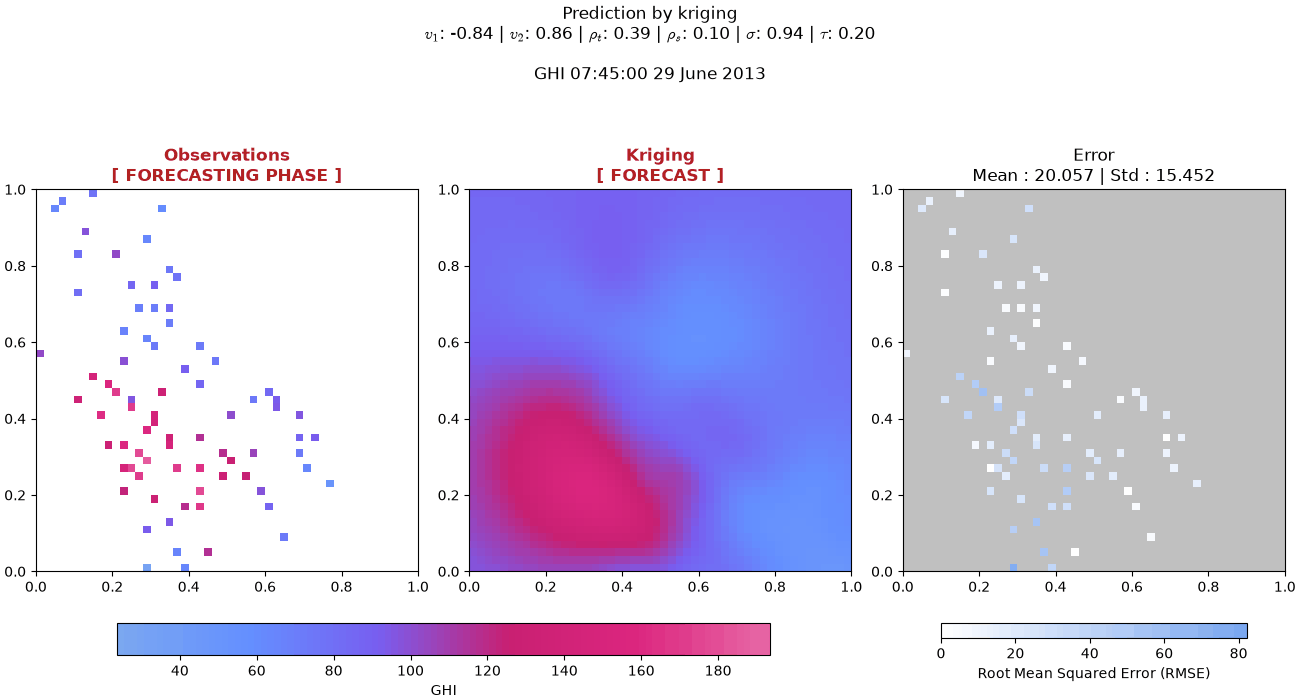}
        \caption{Forecasting at 7:46 on June 29, 2013.}
        \label{fig:3D_krig_forecast}
    \end{subfigure}

    \caption{Sample frames illustrating interpolation and forecasting using kriging on June 29, 2013 (the complete animation is available in the \href{https://github.com/DarenXL/Neural_ST_inference/blob/main/output/results/3D/kriging_animated.gif}{GitHub repository}).}
    \label{fig:3D_krig}
\end{figure}

Using the estimated parameters, we perform spatio-temporal prediction and forecasting.  Figure~\ref{fig:kriging_error_timeseries} illustrates an example over a 30-minute period from 7:21 to 7:50 on June 29, 2013. We suppose that data are available from $t_0 - 19$ (7:21) to $t_0$ (7:40). For these first 20 minutes, spatio-temporal kriging is performed using the parameter vector estimated during the $[t_0 - 19,t_0]$ time window. For the subsequent 10 minutes, to enhance forecast robustness, forecasting is conducted using the sequence of parameter sets from previous time windows, filtered with a half-Gaussian kernel to weight past information based on its recency. Specifically, we evaluate three scenarios: (i) relying on a single lag (i.e., using only the most recently inferred parameter vector), (ii) incorporating up to 8 previous lags weighted by a half-Gaussian kernel with a standard deviation of $2.5$, and (iii) extending to 16 previous lags using a half-Gaussian kernel with a standard deviation of $5$. As shown in Figure~\ref{fig:kriging_error_timeseries}, the kriging error remains non-zero due to inherent observation noise, and the forecasting error naturally increases as predictions move further away from the last available observations. Figure~\ref{fig:3D_krig} displays a frame from both the interpolation and the forecasting. Animations of prediction and forecasting by kriging, along with a conditional simulation, are available in the \href{https://github.com/DarenXL/Neural_ST_inference/blob/main/output/results/3D}{GitHub repository}.

\section{Conclusion} \label{section:conclusion}

To overcome the computational limitations of traditional likelihood-based methods to estimate parameters of spatio-temporal models, we proposed to leverage two existing Simulation-Based Inference approaches: Neural Bayes Estimation (NBE) and Neural Ratio Estimation (NRE). Both methods train a neural network on simulated datasets to efficiently infer parameters from observations at a reduced computational cost, leveraging the advantages of \emph{amortization}. By projecting our data onto a regular grid, we enabled a CNN to operate simultaneously across spatial and temporal dimensions to capture complex, joint spatio-temporal dependencies such as directional advection and non-separability. Our results demonstrate that these approaches successfully estimate a wide range of spatio-temporal models from the Gneiting-Matérn family with advection, even in the presence of substantial missing data.

Transitioning from a two-dimensional (1D space $\times$ 1D time) to a three-dimensional (2D space $\times$ 1D time) scenario is fully supported by our pipeline, but execution times increase drastically when increasing the grid resolution. 
Migrating the simulation phase to GPUs would be an effective way of mitigating this bottleneck and accelerating computation. As for training, considering alternative neural network architectures, such as transformers \citep{vaswani2017attention}, could provide an acceleration without loss of embedding efficiency.
Overcoming these computational challenges is a prerequisite for deploying our method on extremely large scale real-world datasets.

Beyond dimensionality, we would like to challenge our methods on more complex cases. This primarily concerns non-stationary models, where the underlying parameters may vary depending on the position in space or time. While accounting for these variations introduces a much higher degree of flexibility necessary for modeling complex real-world observations, it also increases the dimensionality and computational complexity of the inference process. These challenges will surely provide motivation to further develop our current methods, or to explore alternative SBI approaches.

Another promising direction for future work is extending our framework to Neural Posterior Estimation (NPE). While NBE produces point estimates, NPE leverages recent advances in generative architectures, such as normalizing flows, flow matching, and diffusion models, to directly infer the full posterior distribution in a single forward pass \citep{papamakarios2021normalizing, arruda2025diffusion}. This offers richer inferential insights, including direct access to posterior quantiles and uncertainty quantification. Although training NPE models is computationally more demanding than NBE due to the higher dimensionality of the target distribution, incorporating these expressive density estimators offers significant advantages for applications requiring comprehensive uncertainty assessment.

\section*{Data availability statement}

The data that support the findings of this study are openly available in the \href{https://github.com/DarenXL/Neural_ST_inference}{GitHub repository} of the paper.

\section*{Disclosure statement}

Google Gemini assisted with code development and debugging. The authors reviewed all assisted material and retain full responsibility for the work.

The authors have no relevant financial or non-financial interests to disclose.

\if1\anon{
\section*{Acknowledgements}
This work was supported by the \href{https://chaire-geolearning.org/}{Geolearning chair} of Mines Paris – PSL, funded by Andra, BNP Paribas, France, CCR and the SCOR Foundation for Science.
}
\fi

\bibliography{references}  

\end{document}